\documentclass[12pt,A4]{article}
\usepackage{a4wide,amssymb,latexsym,afterpage,calc,citesort,epsfig,
array,tabularx,amsmath}
\begin{document}
\parindent=0cm
\parskip=1.5mm
\def\bi{\begin{list}{$\bullet$}{\parsep=0.5\baselineskip
      \topsep=\parsep \itemsep=0pt}} \def\ei{\end{list}}
\def\phi{\varphi} \def\-{{\bf --}} \def\vm{v_{max}}
\def\cH{{\cal H}}
\newcommand{\eins}{\mathbf 1}
\begin{center}
  {\LARGE\bf The asymmetric exclusion process:}
\end{center}
\begin{center}
  {\LARGE\bf Comparison of update procedures}
\end{center}
\vskip1.8cm 
\renewcommand{\thefootnote}{\fnsymbol{footnote}}
\setcounter{footnote}{0}
\begin{center}
  {\Large N.\ Rajewsky\footnote{email: \tt nr@thp.uni-koeln.de}, L.\
Santen\footnote{email: \tt santen@thp.uni-koeln.de}, A.\ 
    Schadschneider\footnote{email: \tt as@thp.uni-koeln.de} and} 
\end{center}
\begin{center}
\vskip-0.2cm
  {\Large 
M.\ Schreckenberg\footnote{email: \tt schreck@rs1.uni-duisburg.de} }
\end{center}
\vskip0.6cm
\begin{center}
  $^{1,2,3}$\,Institut f\"ur Theoretische Physik\\ Universit\"at zu K\"oln\\ 
  D--50937 K\"oln, Germany
\end{center}
\begin{center}
  $^{4}$ Theoretische Physik/FB 10\\ 
  Gerhard-Mercator-Universit\"at Duisburg\\ 
  D--47048 Duisburg, Germany\\
\end{center}
\begin{center}
\today
\end{center}
\vskip0.3cm {\large \bf Abstract}\hspace{.4cm} 
The asymmetric exclusion process (ASEP) has attracted a lot of interest 
not only because its many applications, e.g.\ in the context of the kinetics 
of biopolymerization and traffic flow theory, but also because it is a 
paradigmatic model for nonequilibrium systems. Here we study the ASEP for 
different types of updates, namely
random-sequential, sequential, sublattice-parallel and parallel. 
In order to compare the effects of the different update procedures on
the properties of the stationary state, we use large-scale Monte Carlo
simulations and analytical methods, especially the so-called matrix-product
Ansatz (MPA).
We present in detail the exact solution for the model with sublattice-parallel
and sequential updates using the MPA. For the case of parallel update, 
which is important for applications like traffic flow theory,
we determine the phase diagram, the current, and density profiles based
on Monte Carlo simulations. We furthermore suggest a MPA for that case
and derive the corresponding matrix algebra.\\[0.7cm]
\vfill \pagebreak 

\renewcommand{\thefootnote}{\arabic{footnote}}
\setcounter{footnote}{0}
\section{Introduction}

For nonequilibrium systems in low dimensions an understanding can
often be gained by studying rather simple models 
\cite{book,ziabook,spohnbook,ligett,katz,garrido}.
The one-dimensional asymmetric exclusion process (ASEP) has been
used to describe various problems in different fields of interest,
such as the kinetics of biopolymerization \cite{bio} and traffic
\cite{traffic}. On the other hand, the ASEP is so simple that it has
achieved a paradigmatic status for nonequilibrium systems \cite{DErev}. It 
can be mapped onto a surface growth model known as the single-step model
\cite{meakin} and in the appropriate hydrodynamical limit its density 
profile obeys the Burgers equation which is itself
closely related to the KPZ equation \cite{kardar}. The
ASEP can also be viewed as a prototype for so-called boundary-induced
phase transitions \cite{krug}: the boundaries, which can inject and
remove particles from the system, govern -- in a subtle interplay with
the local dynamical rules -- the {\it macroscopic} behaviour of the
model and can produce different phases and phase transitions.

In \cite{derrida92} recursion relations on the system size have been
derived for the ASEP with random-sequential update and open boundary 
conditions. Open boundaries here and in the following
mean that particles are injected at one end of a chain of $L$ sites
with probability $\alpha$ and are removed at the other end with
probability $\beta$. In \cite{derrida92} these recursion relations 
were solved for the special case $\alpha=\beta=1$. This solution
was later extended in \cite{schdom} to general $\alpha$ and $\beta$. 
A very elegant solution of the general case was given at the same time
in \cite{derrida93} using a matrix product Ansatz (MPA) for the weights 
of the stationary configurations similar to the matrix product groundstate 
of certain quantum spin chains \cite{zitt}.
This Ansatz can be used to compute density profiles and 
correlation functions. The relationship between the MPA for quantum spin chains
and one-dimensional models of statistical physics will be discussed
in this paper.

Since then the MPA was extended to find the transient of the model
\cite{stinch}, to describe the ASEP with a defect in form of an
additional particle with a different hopping rate \cite{mallick} or a
blockage \cite{hinsan}, to solve the case of oppositely charged
particles (with hard-core repulsion), which move in opposite
directions (driven by an external electric field) and can interchange
their charge if they meet \cite{evansetal}. The MPA was also used to 
recover solutions of certain integrable reaction-diffusion models 
\cite{resolve}. 

Most of these solutions have been found for random-sequential dynamics. 
In that case the master equation can be 
rewritten as a Schr\"odinger-like equation for a ``Hamiltonian''  with
interactions between nearest-neighbours only \cite{alca}.  
Other updates lead to more complicated master equations with non-local
interactions.

It also turned out that the MPA is not just an Ansatz. The stationary state 
of an one-dimensional stochastic model with arbitrary nearest-neighbor
interactions and random-sequential update can always be written as a
matrix product \cite{sandkrebs}. This is also true for an
ordered-sequential or sublattice-parallel update, which can be shown to be
intimately related \cite{rajsch}. The matrices
in the MPA are generally infinite-dimensional. Therefore, evaluating
physical quantities such as density profiles is still a formidable
task, but often it is at least possible to obtain asymptotic
expressions. However, in certain regions of the parameter space the
matrices can reduce to low-dimensional variables. A simple example is
the case of one-dimensional matrices, which is equivalent to a
mean-field solution.

The implementation of the order of application of the local transition
rates for a given model (the type of update) is an essential part of
the definition of the model, since the transient and even the stationary 
state may differ dramatically \cite{lecaer}. In the following, the ASEP
will be studied with types of update that are often more useful for
Monte Carlo simulations than the random-sequential update.  The aim of
this paper is to use the ASEP as a case study to investigate the
consequences of different types of updates onto the stationary state
of a nonequilibrium model with open boundaries.  Concerning the analytical
treatment, the MPA turns out to be a useful tool.

Hinrichsen \cite{hinri} was the first to apply the MPA to the ASEP 
with sublattice-parallel update and deterministic bulk dynamics. He could 
confirm earlier conjectures for the correlation functions \cite{schuetz2}. 
It is important to note
that this update is substantially different from the fully-parallel
update used for example for modeling traffic flow. Nevertheless, to
our knowledge this is the first model with a discrete-time update which
has been solved using the MPA. The results presented here build partially
on our previous work \cite{work}, where we found a mapping of the ASEP with 
ordered-sequential and sublattice-parallel update onto the random-sequential 
case. For sublattice-parallel update this was done independently by Honecker
and Peschel \cite{honi}. We will explain the mapping in more detail
and give a physical interpretation of the underlying Ansatz. As a
by-product, we solve the model on a ring.

We also study numerically and analytically the ASEP with parallel
update and present new results like the phase diagram and the current.
We show that the model can be described by a matrix-product structure at
least for small systems. The underlying algebra is somewhat different to the
other ones because a third state appears, which is neccessary to
decompose the transfer matrix into a more simple product.

This paper is organized as follows: Section \ref{asepdefanal} gives the
definitions of the model and the updates used and establishes a
precise relationship between the MPA for the stationary state of a
stochastic model and for the groundstate of quantum spin chains.
Section \ref{secnonc} is devoted to the MPA for the ASEP with 
discrete-time updates. In Section \ref{seqsect} the matrix algebra
for the ordered-sequential and the sublattice-parallel update is derived.
In Section \ref{parsect} we propose a MPA for the parallel update by 
mapping the model onto a 3-state model with an ordered-sequential update.
Furthermore, we find a special line in parameter space where the 2-cluster
approximation becomes exact. In Section \ref{compasect} we use the
mathematical results of Section \ref{secnonc} and Monte Carlo simulations
to calculate the current, density profiles and correlation functions for
the discrete-time updates. Finally, Section \ref{discsect}
contains a concluding discussion. We have included several appendices,
which mainly list details of the calculations.

\section{Definition of the ASEP and the MPA}
\label{asepdefanal}

\subsection{The ASEP and the different updates}
\label{asepup}

\setcounter{figure}{0}

Consider sites located on a chain of length $L$. Each site $i$ ($1\le
i \le L$) is either  occupied by a particle ($\tau_i=1$), or it is
empty ($\tau_i=0$). We have three basic mechanisms (see Fig.~1):
{\it Hopping}: We look
at the pair $(i,i+1)$. If we find a particle at site $i$ and 
no particle (``hole'') at site $i+1$, we move the particle one site
to the right with probability $p$. In the remaining cases, nothing happens.
\\ 
{\it Injection} and {\it removal}: At the boundaries particles/holes can
be injected and/or removed. At the left boundary a particle can be injected 
with probability $\alpha$ if site $1$ is empty. At the right end
of the chain a particle will be removed with probability $\beta$. 
\\ 
The model can be generalized to allow hopping in both directions by
introducing a probability $q$ for hopping onto an unoccupied site to the
left. Furthermore, one can also inject particles at the right end and remove 
particles at the left end. These modifications do not change the basic 
features of the model and will not be considered in the following.

We now have to define the {\it order} in
which to perform the hopping, the injection and
the removal in terms of time and space. There 
are four basic ways to do that:
\\ 
(a) {\it random-sequential update}: We pick at random a site $i$. If $2\le i\le
L-1$, each particle has a probability $pdt$ of jumping to the
right (if this site is empty). If $i=1$ , we allow for
hopping to the right with probability $pdt$ and particle
injection with rate $\alpha$ if this site is empty.  
For $i=L$ a particle is removed with rate $\beta$ if the site is 
occupied.
This update is the realization of the usual master equation in continuous 
time. 
A different $p$ would simply result in a rescaling of time (see Sect.\ 
\ref{secnonc}). As a consequence, the phase diagram of
the ASEP with update (a) depends only trivially on $p$. Therefore one can
set $p=1$ which is most efficient for computer simulations.
\\
The following three updates are discrete in time.
\\ 
(b) {\it sublattice-parallel update}: We first use our rule for injection 
(removal) at site
$1$($L$). We then perform our rules for hopping on the
pairs $(2,3)$,$(4,5)$ etc. After that, we update the pairs
$(1,2)$,$(3,4)$ etc ($L$ has to be even). 
This update  can be used efficiently for computer simulations. Its main
advantage for theoretical purposes is that 
its transfer matrix can be written as a product of local terms.
\\ 
(c) {\it ordered-sequential update}: We start at the right end
of the chain and remove a particle at site $i=L$ with probability
$\beta$. We then update the pair $(i=L-1,i=L)$. We continue with pair
$(i=L-2,i=L-1)$ and so forth, until the left end of the chain is reached.
After the update of pair $(i=1,i=2)$, we allow for injection at site
$i=1$. For models where particles only hop to the right this update
may also be called more precisely 'backward-ordered-sequential update'.
\\ 
Obviously, the order of update can also be reversed. For the ASEP, these two 
updates are connected by a particle-hole symmetry: injecting particles
can be regarded as removing holes, and vice versa. Therefore it is
sufficient to study just one of the ordered-sequential updates.
\\ 
We would like to point out that in principle one has to distinguish two 
different types of ordered-sequential updates which one could name
site-ordered-sequential and particle-ordered-sequential, respectively. 
In contrast to the site-ordered-sequential update described in (c) above, 
in the particle-ordered-sequential update the rules
are only applied to {\em occupied} sites, i.e.\ to particles. This might
have a strong effect, as can be seen most easily for the case of small
particle numbers and the completely asymmetric deterministic case $p=1$. 
In the case of site-sequential update a single particle injected at the 
left end moves through the lattice in one timestep (i.e., one sweep 
through the lattice). For the particle-sequential
update a timestep means updating occupied sites only and so the particle
moves only one site. By looking at a lattice with two particles, one can
already see that the two different updates might introduce rather different
correlations. Starting with particles separated by $d$ empty sites, in the 
site-ordered-sequential update the left particle will move to the right until 
it reaches the right particle, which then starts to move. On the other hand,
in the case of particle-ordered-sequential update the particles will stay 
always $d$ or $d-1$ sites apart. For general values of $p$ the situation
is similar.
\\
The differences between these two sequential update procedures manifest
itself also in the solution for periodic boundary conditions. We will come 
back to this point in Section \ref{periodsec}. In the following we will 
always consider the site-ordered-sequential update {\bf --} until stated 
otherwise {\bf --} to which we will refer as ordered-sequential update for 
brevity.
\\
(d) {\it parallel update}: The rules for hopping, injection, and removal 
are applied simultaneously to all sites of the whole chain\footnote{Note
that for the parallel update the partially symmetric ASEP (hopping to right 
and left with probabilities $p$ and $q$, respectively) might lead to 
ambiguities. One therefore has to set $q=0$.}.
\\ 
The parallel update usually produces the strongest
correlations and is used for traffic simulations \cite{traffic}. In the
case of the ASEP, it is nearly identical to the 
particle-ordered-sequential update.

Fig.\ 2 illustrates the updates (b) and (c).

If the ASEP with updates (a)-(c) is put on a ring (no
injection/removing of particles and periodic boundary conditions), a
trivial stationary state (where correlations are absent, see below) is
reached, while update (d) produces a particle-hole attraction \cite{schreck}.
This already shows that different updates might yield a different behaviour.

For analytic calculations usually the random-sequential update is most
convenient since it can be formulated in terms of a ``Hamiltonian'' with
nearest-neighbour interaction. In Monte Carlo simulations, however,
ordered-sequential updates can be implemented more effectively.

At this point it is necessary to point out the existence of some
confusion in the nomenclature of the different updates. The random-sequential 
update (a) is sometimes simply called 'sequential update', as is the
ordered-sequential one. In several publications the sublattice-parallel
update (b) is called 'parallel' which makes it necessary to refer
to the parallel update (d) as 'fully-parallel'. We urge the reader to
check carefully which type of update is actually used when consulting
the literature. In order to avoid further confusion we will use 
the terminology which seems to us the most precise.

\subsection{Master equation and quantum formalism}
Our starting point is the master equation for an arbitrary one-dimensional 
stochastic process. Following \cite{alca} we rewrite this equation as a 
Schr\"odinger-like equation in imaginary time. We consider a chain of 
$L$ sites $j$ with state variables $\tau_j=0,1$. Generalizations to the 
case where the state variables can take more than two values
 are straightforward.
A configuration of the whole system will be denoted by $\{ \tau\}=\{
\tau_1, \tau_2, ..., \tau_L\}$, its weight by $P(\{ \tau\},t)$.
 
The master equation then has the form
\begin{equation}
\frac{\partial}{\partial t} P(\{ \tau \},t)=\sum_{\{\tau' \}}
\left[ w(\tau'\rightarrow \tau )P(\{\tau' \},t) 
- w(\tau \rightarrow \tau') P(\{ \tau \},t)\right]  
\end{equation}
where $w(\tau \rightarrow \tau')$ denotes the rate for a transition
from $\{ \tau\} $ to $\{ \tau'\}$.\\
Let us now define a ket state $|P(t)\rangle$ in the following way.
We take an orthonormal basis in the configuration space $\{\tau \}$,
\begin{equation}
|\tau \rangle=|\tau_1,...,\tau_L \rangle
\end{equation}
with $\langle\tau'|\tau\rangle=\delta_{\tau', \tau }$
and define 
\begin{equation}
|P(t)\rangle=\sum_{\{\tau\}} P(\{\tau\},t) |\tau\rangle \, ,
\end{equation}
i.e.\ we have $P(\{\tau\},t)=\langle\tau |P(t)\rangle$.
It is then easy to see that 
\begin{equation}
\label{schroed}
\frac{\partial}{\partial t} |P(t)\rangle =-\cH \, |P(t)\rangle 
\end{equation}
holds, with the (generally non-hermitean) ``Hamiltonian'' $\cH$ given by
\begin{equation}
\label{hamm1}
\langle \tau|\cH|\tau'\rangle=-w(\tau ' \rightarrow \tau)
\end{equation}
for the off-diagonal elements ($\tau\neq\tau'$) and by
\begin{equation}
\label{hamm2}
\langle \tau|\cH|\tau\rangle=\sum_{\{ \tau \} \ne \{ \tau'
  \}}w(\tau \rightarrow \tau')
\end{equation}
for the diagonal elements. 
\\
From (\ref{schroed}) one can see that the stationary state $|P_0\rangle$ 
of the stochastic model corresponds to the ``groundstate''
with ''groundstate energy'' zero ($E_0=0$) of the ''quantum spin chain'' 
defined by (\ref{hamm1}), (\ref{hamm2}), i.e.\ 
\begin{equation}
\label{GZeq}
\cH|P_0\rangle = 0.
\end{equation}

The conditions (\ref{hamm1}), (\ref{hamm2}) guarantee that the real
parts of the other eigenvalues $E_\lambda$ of $\cH$ are non-negative. 
Using the bra groundstate $\langle 0|=\sum_{\{\tau\}}\langle\tau|$, the
average of an observable ${\cal A}(\{\tau\})$ at time $t$ is given by
\begin{equation}
\langle {\cal A} \rangle(t) = \sum_{\{\tau\}} {\cal A}(\{\tau\})
P(\{\tau\},t)=\langle 0|{\cal A}|P(t)\rangle.
\end{equation}
Expanding the initial state $|P(t=0)\rangle =\sum_\lambda a_\lambda
|\psi_\lambda\rangle$ in terms of the eigenkets $|\psi_\lambda\rangle$ 
with eigenvalues $E_\lambda$ of $\cH$ this can be rewritten as
\begin{equation}
\label{time}
\langle {\cal A} \rangle(t)=\langle 0|{\cal A}|P(t)\rangle
=\langle 0|{\cal A}e^{-\cH t}|P(t=0)\rangle
=\sum_\lambda a_\lambda e^{-E_\lambda t}
\langle 0|{\cal A}|\psi_\lambda\rangle.
\end{equation}
This shows that the behaviour for large times $t$ is governed by
the low-lying excitations.
\\
We now restrict ourselves to stochastic processes with random-sequential
dynamics and local and homogenous 
transition rates $\Gamma_{\tau_j \tau_{j+1}}^{\tau'_j \tau'_{j+1}}$ 
(denoting the rate for a local transition of sites $j$ and $j+1$ from 
(${\tau'_j \tau'_{j+1}}$) to (${\tau_j \tau_{j+1}}$)),
which do not depend on time.  In this case $\cH$ is a sum of local
Hamiltonians with nearest-neighbour interaction only \cite{alca}:
\begin{equation}
\label{hamml2}
\cH=\sum_{j=1}^{L}h_j \quad \,
\end{equation}
with
\begin{equation}
h_j= 
\begin{pmatrix} 
\Gamma_{01}^{00}+ \Gamma_{10}^{00} +\Gamma_{11}^{00} &
-\Gamma_{00}^{01} & -\Gamma_{00}^{10} & -\Gamma_{00}^{11} \\
-\Gamma_{01}^{00} & \Gamma_{00}^{01}+ \Gamma_{10}^{01} +\Gamma_{11}^{01} &
-\Gamma_{01}^{10} & -\Gamma_{01}^{11} \\
-\Gamma_{10}^{00} & -\Gamma_{10}^{01} & \Gamma_{00}^{10}+
\Gamma_{01}^{10} +\Gamma_{11}^{10} &  -\Gamma_{10}^{11} \\
-\Gamma_{11}^{00} & -\Gamma_{11}^{01} & -\Gamma_{11}^{10} & \Gamma_{00}^{11}+ \Gamma_{01}^{01} +\Gamma_{10}^{11} 
\end{pmatrix}
\end{equation}
(the basis is $(00)$,$(01)$,$(10)$,$(11)$). Each column of $h_j$ adds
up to $0$, because the probability has to be conserved.\\
For the updates (b)-(d) it is generally not possible
to write $\cH$ in the form (\ref{hamml2}) since one would have
to include terms acting on sites which are not neighboured. It
is then more suitable to use directly the transfer matrix $T$ describing
the update of the whole chain during one timestep. In
the case of the ordered-sequential update, $T$ is simply a product of
the local transfer matrices $t_j=h_j-1$. It follows
\begin{equation}
|P(t+1)\rangle =T |P(t)\rangle,
\end{equation}
which means that in order to find the stationary state, one
has to solve for the eigenvector of $T$ with eigenvalue $1$.

\subsection{Optimum groundstates and MPA for quantum spin chains}
\label{sectoptimum}

The construction of optimum ground states for quantum spin chains
via matrix products was introduced in \cite{zitt} (see also \cite{hakim,FNW} 
and \cite{nigge} for further references). 
Let us consider a Hamiltonian for a quantum spin chain (with
periodic boundary conditions) of the form 
$\cH'\, =\, \sum_{j=1}^{L}h_j', $
where $h_j'$ is the local {\it hermitian} Hamiltonian and independent of 
$j$, acting only on spin $j$ and $j+1$.
\\
It is always possible to set the lowest eigenvalue
of $h_j'$ equal to zero by adding a suitable constant. Then
$h_j'$ is positive-semidefinite and since $\cH'$ is the sum of
positive-semidefinite operators, it follows that zero is a lower bound 
for the ground state energy $E_0$ of $\cH'$, i.e.\ $E_0\geq 0$.
Usually, $E_0$ is greater than zero ($E_0>0$) and the global groundstate 
involves also excited states of $h_j'$. Therefore, a construction of the
global groundstate is usually very difficult.
\\
However, there are special cases where $E_0$ is equal to zero, 
\begin{equation}
\cH'|\psi_0\rangle \, =\, 0 
\end{equation}
and therefore for all $j$
\begin{equation}
\label{optimal}
h_j' \, |\psi_0 \rangle \, = \, 0.
\end{equation}
A state $|\psi_0\rangle$ is called {\it optimum groundstate} of $\cH'$ 
if and only if condition (\ref{optimal}) holds. This implies that
the groundstate energy is independent of the system size, i.e.\ there
are no finite-size corrections.\\
The idea is now to construct groundstates by means
of a product of matrices,
\begin{equation}
\label{mpg}
|\psi_0\rangle\, = \, {\rm Tr}\,( m_1 \otimes m_2 \otimes ... \otimes m_L )\, ,
\end{equation}
where the entries of matrix $m_j$ are spin-1 single-site states 
and the symbol $\otimes$ denotes the usual matrix multiplication
of matrices with a tensor product of the matrix elements. Note that
$m_1 \otimes m_2 \otimes ... \otimes m_L $ is still of the same size
as the matrices $m_j$, but its elements are large linear combinations of
tensor product states.
The trace assures the translation invariance of the groundstate.
For non-periodic boundary conditions it has to be replaced by a suitable
linear combination of the elements of 
$m_1 \otimes m_2 \otimes ... \otimes m_L $.\\
As an example \cite{zitt}, the 
ground state of a large class of antiferomagnetic
spin-1 chains can be constructed using the $S^z$ eigenstates
$|0\rangle_j$ and $|\pm \rangle_j$ by
\begin{equation}
\label{machtrix} 
m_j =\begin{pmatrix} 
a \, |0\rangle_j & b \, |+\rangle_j \\
c \, |-\rangle_j & d \, |0\rangle_j
\end{pmatrix} \, ,
\end{equation}
where the $a,b,c,d$ are real numbers. Condition (\ref{optimal}) requires
\begin{equation}
\label{vernicht}
  h_j'\, (m_j\otimes m_{j+1}) \, = \, 0 \quad ,
\end{equation} 
i.e.\ all four elements of $m_j\otimes m_{j+1}$ are local groundstates
of $h_j$. Let us now write
\begin{equation}
m_j=A_0 \cdot |0\rangle + A_- \cdot |-\rangle + A_+ \cdot |+\rangle \,  
\end{equation}
with suitable $2\times 2$ matrices $A_0$ and $A_{\pm}$. The
'$\cdot$' denotes a product of each entry of the matrix to the left
with the single-site state on the right.
\\
It is obvious that (\ref{mpg}) can be written
equivalently as
\begin{equation}
\label{stoch} 
|\psi_0 \rangle ={\rm Tr} 
\bigg[\bigg( \begin{array}{c} A_0 \\ A_- \\ A_+ \end{array}\bigg)
\otimes\bigg( \begin{array}{c} A_0 \\ A_- \\ A_+ \end{array}\bigg)
\otimes ... \otimes\bigg( 
\begin{array}{c} A_0 \\ A_- \\ A_+ \end{array}\bigg)\bigg]\, ,
\end{equation}
or, defining a column vector
\begin{equation}
\label{column}
A\, = \, \bigg( \begin{array}{c} A_0 \\ A_- \\ A_+ \end{array}\bigg)\quad ,
\end{equation}
as
\begin{equation}
\label{mpg2}
|\psi_0 \rangle ={\rm Tr} (A\otimes A \otimes ... \otimes A) \quad .
\end{equation}
The condition (\ref{vernicht}) then be rewritten as
\begin{equation}
\label{vernichtA}
 h_j'\, (A\otimes A) \, = \, 0 \quad .
\end{equation}

This means that there are two equivalent ways to write $|\psi_0\rangle$.
While (\ref{mpg}) uses a product of  matrices with vectors as
entries (the usual notation for quantum spin chains), (\ref{mpg2}) 
expresses $|\psi_0\rangle$ as  a product of vectors with matrices as entries.
The original idea of Derrida et.al.\ \cite{derrida93} was to
construct the stationary state $|P_0\rangle$ of a stochastic process
defined by (\ref{hamml2}) as a suitable linear form of a 
product of matrices where each matrix corresponds to a single-site
state precisely as in (\ref{mpg2}).
\subsection{MPA for stochastic systems}

In the following we want to describe how the MPA can be applied to 
stationary states $|P_0\rangle$ of stochastic systems. In contrast to 
the case of quantum spin chains we already know the corresponding
``groundstate energy'' of the stochastic Hamiltonian $\cH$
defined by (\ref{hamm1}) and (\ref{hamm2}). From (\ref{schroed}) 
we see that $\cH|P_0\rangle = 0$ and therefore the groundstate energy
is zero, independent of the system size. This hints at the applicability 
of the MPA and a possible generalization of the optimum groundstate concept.

Indeed it turns out that, using an MPA of the form (\ref{mpg2}),
(\ref{optimal}) can be replaced by a more general condition by allowing 
for a divergence-like term \cite{resolve}, i.e.\
\begin{equation}
\label{cancel}
h_j (A\otimes A)={\bar A} \otimes A - A \otimes {\bar A} \ ,
\end{equation}
where the vector $\bar{A}$ can be different from the vector $A$.
It is easy to see that (for periodic boundary conditions) the divergence-like
terms cancel after summing over $j$. Hence, the stationary state of the
stochastic process described by (\ref{hamm1}) and (\ref{hamm2}) 
is of the form (\ref{mpg2}).

For hermitian Hamiltonians one always has $\bar A = A$ and (\ref{cancel}) 
reduces to (\ref{vernichtA}). Therefore one can regard (\ref{cancel}) as the
generalization of the optimum groundstate concept to non-hermitian 
Hamiltonians.

Again it is possible to generalize these ideas to treat non-periodic
boundary conditions. As an example we briefly review the solution 
of the ASEP with random-sequential update and open boundary conditions. 
The ``Hamiltonian'' reads (hopping rate $p$, feeding and removal
rates $\alpha$,$\beta$):
\begin{equation}
\label{hasep1}
H={\bar h_1}+{\bar h_L}+\sum_{j=1}^{L-1}h_j 
\end{equation}
with the boundary terms
\begin{equation}
\label{hasep2}
{\bar h_1}= \begin{pmatrix} \alpha & 0 \\ -\alpha & 0 \\ \end{pmatrix}\ ,
\qquad
{\bar h_L}= \begin{pmatrix} 0 & -\beta \\ 0 & \beta \\ \end{pmatrix}
\end{equation}
and the bulk ``Hamiltonian''
\begin{equation}
\label{hasep3}
h_j=\begin{pmatrix} 0 & 0 & 0 & 0 \\ 0 & 0 & -p & 0 \\ 0 & 0 & p & 0
  \\ 0 & 0 & 0 & 0 \end{pmatrix} \quad .
\end{equation}
From (\ref{time}) we can see that $p$ only rescales time (and $\alpha$
and $\beta$) and that it would therefore be sufficient to study $p=1$.

Following the MPA, the matrix for the
particle (hole) is denoted by $D$ ($E$), so that
$A=\left(\begin{array}{c} E \\ D \end{array}\right)$. 
Since the ASEP with open
boundary conditions has (in general) no translational invariance,
the trace in (\ref{mpg2}) is replaced with a scalar product:
\begin{equation}
\label{mpg3}
|P_0\rangle= \frac{1}{Z_L}\langle\langle W|A\otimes \cdots\otimes 
A|V\rangle\rangle \, ,
\end{equation}
where the
normalization constant $Z_L$  is equal to $Z_L=\langle W|C^{L}|V\rangle$
with $C=E+D$.
The brackets $\langle \langle ... \rangle \rangle$ indicate that the 
scalar product 
is taken in each entry of the vector $A\otimes ... \otimes A$.
\\
More explicitly (\ref{mpg3}) means that the weight $P(\tau_1,\ldots,
\tau_L)$ of a configuration $\{\tau\}$ in the stationary state
is given by
\begin{equation}
\label{mpgweights}
P(\tau_1,\ldots,\tau_L)=\frac{1}{Z_L}\langle W|\prod_{j=1}^L\left[
\tau_jD+(1-\tau_j)E\right]|V\rangle.
\end{equation}
Thus one has a simple recipe for the calculation of the (unnormalized)
weight of an
arbitrary configuration $\{\tau\}$: Translate the configuration into
a product of matrices by identifying an empty site ($\tau_j=0$) with $E$
and an occupied site ($\tau_j=1$) with $D$. For example, the configuration
$011001\cdots$ corresponds to the product $EDDEED\cdots=ED^2E^2D\cdots$.
Then multiply by the vectors $\langle w|$, $|v\rangle$ from the left and
right, respectively.

We assume that $\bar{A}$ in (\ref{cancel}) is of the form $\bar{A}=\left(
\begin{array}{c} \bar{E} \\ \bar{D} \end{array}\right)$
where $\bar E$ and $\bar D$ denote matrices acting in the same vectorspace
as $E$ and $D$.
Executing the sum (\ref{hamml2}) in (\ref{cancel}) leads to a cancelation 
of all terms in the bulk of the chain. The remaining terms
at the boundaries vanish if the  vectors $\langle W|$ and $|V\rangle$ 
are chosen appropriately:
\begin{equation}
\label{bound}
\langle W|{\bar h_1} A =-\langle W|{\bar A} \quad , \quad 
{\bar h_L} A|V\rangle= {\bar A}|V\rangle \, .
\end{equation}
Inserting (\ref{hasep2}) and (\ref{hasep3}) into (\ref{cancel}),
(\ref{bound}) we
get a system of quadratic equations in $E,D,{\bar E}$ and ${\bar D}$
which is called the {\it algebra} of Ansatz (\ref{cancel}).
The dimension of the matrices is not determined by the
Ansatz. However, taking ${\bar E}=\eins =-{\bar D}$, the equations
reduce to the ``DEHP algebra'' \cite{derrida93}
\begin{eqnarray}
p\, DE & = & D+E \quad ,\label{derr1}\\
\alpha \langle W| E & = & \langle W| \quad ,\label{derr2} \\ \beta
D|V\rangle & = & |V\rangle\quad .\label{derr3}
\end{eqnarray} 
Equation (\ref{derr1}) can be guessed intuitively: 
the current $J_L(j)$ (describing the flux of particles through 
bond $j$ of a chain of length $L$) has to be constant throughout the chain.
$J_L(j)$ is given by \cite{derrida93}
\begin{equation}
\label{current}
J_L(j)=\frac{p}{Z_L}\langle W|C^{j-1}DEC^{L-j-1}|V\rangle \quad ,
\end{equation}
and we see that $DE\propto C=E+D$ is the simplest way to achieve a constant 
current.
\\
It is possible to derive explicit expressions for $D$, $E$, $\langle W|$
and $|V\rangle$ \cite{derrida93}. It turns out, that one finds
one-dimensional representations if and only if
\begin{equation}
\alpha+\beta=p.
\end{equation}
In all the other cases, the matrices are infinite-dimensional.
\\
Up to this point, the MPA appears to be just an Ansatz for the stationary 
state. However, it can be shown \cite{sandkrebs} that 
the MPA (\ref{mpg3}) with the ``cancellation-mechanism''
(\ref{cancel}) is an {\it equivalent} reformulation of the
master equation for a stochastic process with random-sequential
update and nearest-neighbour interaction. Therefore it is of general 
interest to study the
quadratic algebras which are produced by (\ref{cancel}) and to
try to find explicit representations \cite{essler,rittenberg,mallicksandow}.
\\
In the next section, we will construct the stationary state
for the ASEP with sequential and sublattice parallel update. Note
that we do not have a ``Hamiltonian'' of the form (\ref{hamml2})
in this case and that the ``cancellation mechanism'' will not be appropriate.

\section{MPA for the ASEP with updates in discrete time} 
\label{secnonc}

In this section we will generalize the results for the random-sequential
update to discrete-time updates. In Section \ref{seqsect} we
solve the ASEP for sublattice-parallel and sequential updates using
the MPA. In Section \ref{parsect} we conjecture a MPA for 
the parallel update.

Let us briefly discuss the precise connection between the random-sequential 
update and the ordered-sequential update, say from the
right to the left, for an arbitray one-dimensional stochastic process
with nearest neighbor interaction. We have seen, that in this case
the "Hamiltonian" describing the random-sequential dynamics is of the 
form (\ref{hamml2}), $\cH=\sum_{j=1}^{L} h_j$, while for the 
ordered-sequential udpate one has to use the transfer matrix 
\begin{equation}
\label{trans}
T=\prod_{j=1}^{L} t_j
\end{equation}
with local matrices $t_j$ connecting the sites $j$ and $j+1$.
From (\ref{hamml2}) and (\ref{time}) it is clear that one of the 
parameters can be used to rescale the time unit; in the case of the 
ASEP this leads to expressions which are independent of the hopping 
probability $p$. For the update (\ref{trans}) with  discrete time, 
this is obviously impossible and we expect the phase diagram of the ASEP 
to depend on $p$ nontrivially. Only in the limit of vanishing densities 
these two updates can be mapped onto each other \cite{honi} by inserting 
$t_j=h_j-\eins$ (where $\eins$ denotes the identity matrix) into 
(\ref{trans}) and expanding the product. For nonvanishing
densities, the equations (\ref{hamml2}) and (\ref{trans})
are connected nontrivially via additional, non-locals terms.

\subsection{Sublattice-parallel and ordered-sequential update}
\label{seqsect}

We now solve the ASEP with ordered-sequential (update (c)) and
sublattice-parallel (update (b)) dynamics. So far, the ASEP with the
latter update and deterministic hopping has been studied by Sch{\"u}tz
\cite{schuetz2} and Hinrichsen \cite{hinri}. A brief account of
our work has been given in \cite{work}.
For simplicity, we will first concentrate on the ordered-sequential 
update (c) from the right to the left. Since
this update is discrete in time, a stochastic ``Hamiltonian''
of the form (\ref{hamml2}) is not at hand. Therefore,
the transfer matrix $T_{\leftarrow}$ has to be used. This means
by definition
\begin{equation}
|P(t+1)\rangle \, =\, T_{\leftarrow}|P(t)\rangle
\end{equation}
The stationary state 
$|P_0\rangle$ must not change under the action of the
$T_{\leftarrow}$ and therefore is eigenvector with eigenvalue $1$ of 
$T_{\leftarrow}$:
\begin{equation}
|P_0\rangle\, =\, T_{\leftarrow} |P_0\rangle\, .
\label{ordseqstat}
\end{equation}
Let us now
explicitly write down $T_{\leftarrow}$. The boundary conditions can be 
represented by operators ${\cal R}$ and ${\cal L}$ acting on site $j=L$ 
and $j=1$, respectively:
\begin{equation}
{\cal R} \;=\; \left( \begin{array}{cc}
  1 & \beta \\ 0 & 1-\beta \end{array} \right) \,,
\hspace{10mm}
{\cal L} \;=\; \left( \begin{array}{cc}
  1-\alpha & 0 \\ \alpha & 1 \end{array} \right)\,.
\end{equation}
The basis chosen for ${\cal R}$ and ${\cal L}$ is $(0,1)$. The
update-rule for any pair of sites $(j,j+1)$ can be written as
\begin{equation}
\label{Interaction}
{\cal T}_{j} \;=\; \left(
\begin{array}{cccc}
  1 & 0 & 0 & 0 \\ 0 & 1 & p & 0 \\ 0 & 0 & 1-p & 0 \\ 0 & 0 & 0 & 1
\end{array} \right)\,.
\end{equation}
The basis is $(00,01,10,11)$, and we have 
\begin{equation}
  T_\leftarrow=L\cdot T_1\cdot ... \cdot T_{L-1}\cdot R
\end{equation}
with
\begin{eqnarray}
  L &=& {\cal L}\otimes\eins\otimes ... \otimes\eins\,,\nonumber\\ 
  R &=&   \eins\otimes ... \otimes\eins\otimes{\cal R}\,,\\ 
  T_j &=& \eins\otimes\eins ... \otimes{\cal T}_{j}\otimes\eins ...
  \otimes\eins\,, \nonumber
\end{eqnarray}
where $\eins$ denotes the identity matrix.
\\
Formally, a ``Hamiltonian'' can be defined as
$\cH_{\leftarrow}=\eins-T_{\leftarrow}$. However,
$\cH_{\leftarrow}$ cannot be written simply as a sum of {\it local}
``Hamiltonians''. This means that while we can try a MPG Ansatz (\ref{mpg3})
we cannot use the ``cancellation''-mechanism (\ref{cancel}).\\
However, the sequential nature of $T_{\leftarrow}$ suggests
another mechanism:
\def\edmatrix{\Bigl(\hspace{-1.5mm}
    \begin{array}{c} E \\[-1mm] D
    \end{array}\hspace{-1.5mm} \Bigr)}
\def\abmatrix{\Bigl(\hspace{-1.5mm}
    \begin{array}{c} \hat{E} \\[-1mm] \hat{D}
    \end{array}\hspace{-1.5mm} \Bigr)}
\begin{equation}
\label{Equations}
{\cal T} \, \left[ A \otimes \hat{A} \right] \;=\;
\hat{A}\otimes A \,,
\end{equation}
\begin{equation}
\label{boundEquat}
\langle W| {\cal L} \hat{A} \;=\; \langle W| A\,,
\hspace{15mm}
{\cal R} A |V\rangle\;=\; \hat{A}|V\rangle \,,
\end{equation}
where $\hat{A}=\left(\begin{array}{c} \hat{E} \\ \hat{D} \end{array}\right)$
with square matrices $\hat{E},\hat{D}$. \\
This means that a ``defect'' $\hat{A}$ -- corresponding to a local
perturbation of the stationary state defined by (\ref{ordseqstat}) -- is 
created in the beginning of an update at site $j=L$, which
is then transported through the chain, until it reaches the left end
and disappears.
\\ 
Equation (\ref{Equations}) leads to the following bulk algebra:
\begin{eqnarray}
\label{BulkAlgebra}
[E,\hat{E}] =[D,\hat{D}] &=& 0 ,\nonumber \\
E\hat{D}+pD\hat{E} &=& \hat{E}D ,\\
(1-p)D\hat{E} &=& \hat{D}E ,\nonumber
\end{eqnarray}
and the boundary conditions
\begin{equation}
\label{BoundaryConditions}
\begin{array}{c}
   \langle W| \hat{E} (1-\alpha) \;=\; \langle W| E\, , \\[1mm]
   \langle W| (\alpha \hat{E}+\hat{D}) \;=\; \langle W| D \, ,
\end{array}
\hspace{2mm}
\begin{array}{c}
   (1-\beta)D|V\rangle \;=\; \hat{D}|V\rangle, \\[1mm]
   (E+\beta D)|V\rangle \;=\; \hat{E}|V\rangle .\end{array}
\end{equation}

The ordered-sequential update in the opposite direction (left to right)
can be treated in the same way. The stationary state is given by
\begin{equation}
\label{statright}
|P_0\rangle_\rightarrow= \frac{1}{Z_L}\langle\langle W|\hat{A}\otimes 
\cdots \otimes \hat{A}|V\rangle\rangle
\end{equation}
with the same mechanism (\ref{Equations}) and (\ref{boundEquat}).
However, it is more convenient to use the particle-hole symmetry for
the calculation of averages.

For completeness let us briefly discuss the sublattice-parallel
update (b). In this case the transfer matrix has the structure 
\cite{hinri,honi} 
\begin{equation}
T_{s-\parallel}=T_1\cdot T_2
\end{equation}
with
\begin{eqnarray}
T_1&=&{\cal L}\otimes{\cal T}\otimes{\cal T}\cdots\otimes{\cal T}
\otimes{\cal R},\nonumber\\
T_2&=&\phantom{{\cal L}}{\cal T}\otimes{\cal T}\cdots\otimes{\cal T} 
\otimes{\cal T},
\end{eqnarray}
and the MPA for the stationary state is of the form
\begin{equation}
|P_0\rangle_{s-\parallel}= \frac{1}{Z_L}\langle\langle W|\hat{A}\otimes 
A\otimes\hat{A}\otimes A \cdots \otimes \hat{A}\otimes A|V\rangle\rangle.
\end{equation}
One can now use exactly the same mechanism (\ref{Equations}), 
(\ref{boundEquat}) as in the ordered-sequential case and thus obtains the
same algebra (\ref{BulkAlgebra}), (\ref{BoundaryConditions}).
For $p=1$ the algebra was first derived and solved by Hinrichsen 
\cite{hinri} using a two-dimensional representation for  $E$, $D$,
$\hat{E}$, $\hat{D}$, $\langle W|$, $|V\rangle$. 

The fact that the ordered-sequential and the sublattice-parallel
update lead to the same algebra (\ref{BulkAlgebra}), 
(\ref{BoundaryConditions}) implies the existence of an intimate relationship
between the averages of observables. Although the stationary states
themselves are different, they are connected via transformations, and it 
can be shown that the density profile of
the ordered-sequential update from the left (right) to the right (left)
corresponds to the density of the even (odd) sites produced
by the sublattice-parallel update \cite{rajsch}. This result holds
for arbitrary stochastic models with nearest neighbor interactions.
\\
One can check that for 
\begin{equation}
\label{mean}
(1-\alpha)(1-\beta)=1-p\quad ,
\end{equation}
a one-dimensional solution of the algebra exists\footnote{If we allow
hopping in both directions, this line is given by \cite{work} 
$(1-\alpha)(1-\beta)(1-q)=1-p$ where $q$ is the hopping probability 
to an empty site on the left.}. This equation defines a line in the 
phase diagram where the mean-field solution becomes exact. It 
turns out that this line touches all phases. This makes it possible to
calculate quantitites such as the current 
very easily, because the analytic expression for the current
does not change inside a phase.

For general values of $\alpha$, $\beta$ and $p$ the algebra 
(\ref{BulkAlgebra}) and (\ref{BoundaryConditions}) can be mapped 
onto the generalized DEHP-algebra \cite{work}. 
This is shown in App.\ \ref{appmapp} and will be used later
in Section \ref{secphasetrans}. App.\ \ref{appsymm} discusses the 
consequences of the particle-hole symmetry for the two ordered-sequential
updates in more detail. App.\ \ref{appsymmdiff} deals with the special case
of symmetric diffusion.

\subsection{Parallel update}
\label{parsect}

As far as analytic approaches are concerned, this update 
poses the greatest difficulties, since
it produces the strongest correlations.
In the case of the ASEP, this becomes obvious when
looking  at the model put on ring: All updates, except the parallel 
\cite{schreck} and particle-sequential \cite{evansneu} one, lead to a 
trivial state where correlations are absent. In the case of the parallel 
update, it is known \cite{schreck} that a {\it particle-hole attraction} 
appears: the probability to find an empty site in front of 
an occupied site is enhanced compared to the mean-field result. 
\\
The model with parallel update and periodic boundary conditions has first 
been solved exactly in \cite{schreck,traffic} 
using a cluster approximation (see below). Here the weights are decomposed 
into products of pairs of sites overlapping just one site. It turned out 
that this 2-cluster approximation becomes exact for periodic 
boundary conditions.
\\
We now make an analogous calculation for the model with open boundary
conditions. The goal is to find the parameter set for which the
density profile is flat. Note that a flat density profile does not
necessarily mean that there are no correlations between the sites. 
We denote the probability for the pair configuration $\tau_i \tau_{i+1}$ 
at site $i,i+1$ by $P_{\tau_i \tau_{i+1}}$($\tau_i=1,0$ and $i=1,...,L$). 
Let us assume that such a probability for a certain
pair configuration is independent of the position of the pair. The
condition $P_{01}=P_{10}$  leads to a flat density profile
($\langle \tau_1\rangle=\langle \tau_{2}\rangle=...=\langle \tau_L\rangle$), 
but $P_{\tau_i \tau_{i+1}}$ is not necessarily equal to 
$P_{\tau_i}\cdot P_{\tau_{i+1}}$.
\\
The 2-cluster approximation corresponds to a factorization of the
weight $P(\tau_1, \tau_2, ..., \tau_L)$:
\begin{equation}
\label{cluster}
P(\tau_1, \tau_2, ..., \tau_L)=R_{\tau_1}\cdot P_{\tau_1\tau_2}\cdot
P_{\tau_2\tau_3}\cdot ...\cdot P_{\tau_{L-1}\tau_L}\cdot
R_{\tau_L}\quad ,
\end{equation}
where the $R$'s reflect the influence of the boundaries; they can be set
to $R_0=1$ and $R_1=r$.
\\
It is sufficient to study a system of 3 sites. The generalization to 
larger systems is straightforward \cite{schreck}.
\\ 
The (exact) master equation for the stationary state reads
$x=Tx $
where $x$ is a vector containing the 8 possible configurations, and
$T$ is the $8\times 8$-dimensional transfer matrix for $L=3$. By 
inserting Ansatz (\ref{cluster}) into this master equation
it is straightforward to show that (\ref{cluster}) is exact if and only if
\begin{equation}
\label{meanfield}
(1-\alpha)(1-\beta)=1-p.
\end{equation}
This means that the condition we have found for a constant
density profile is exactly the same as for the 
other discrete-time updates, see (\ref{mean}). The reason for 
this coincidence is still unkown. Note that (\ref{meanfield}) is 
{\it not} a simple mean-field line. 

In the following we will propose a MPA for the parallel update.
(\ref{cluster}) suggests that
matrices denoting pairs of particles and/or holes should be used.
The main problem is that the resulting algebra will be very
complex, since it will contain many variables (matrices). 
On the other hand, if we use a MPA of the form (\ref{mpg3}), a simple
mean-field solution will not be found, e.g. there 
will be no scalar solution for the algebra. However, such an Ansatz
could produce (\ref{cluster}) under condition (\ref{meanfield}) in the
form of a low-dimensional representation. 
\\
Furthermore, even if a choice for the MPG is made, we still have
to find a mechanism which ensures stationarity. The transfer 
matrix, however, cannot be decomposed easily in products or sums of
local terms.
\\
The main difference between the parallel update and the ordered-sequential 
update from the left to the right is that in the parallel update, 
a particle can move only {\em one} site to the right (per update of the chain).
This enables us to use the ordered-sequential update, if 
we introduce a third state for particles that have been moved.
The local (sequential) update operator 
then has to transform a third-state particle 
into a ``normal'' particle  in the following update step.
\\
We write down the Ansatz
\begin{equation}
  |P_0\rangle \;=\;{Z_L}^{-1}\; \langle W| \, \left( \hspace{-1.5mm}
    \begin{array}{c} E \\[-1mm] D \end{array} \hspace{-1.5mm} \right)
  ^{\otimes L} |V\rangle\,\, ,
\end{equation}
which has to satisfy
\begin{equation}
\label{masterpara}
  T_{\parallel} \, |P_0\rangle = |P_0\rangle\quad .
\end{equation}
The update-rule ${\cal T}$ for any pair of sites $(i,i+1)$ is now
nine-dimensional. However, since the third state must not appear {\it after}
the update, the last four rows and every third column are irrelevant
(here set to zero). The explicit expression for ${\cal T}$ can 
be found in Appendix \ref{app3state}.
We have 
\begin{equation}
  T_{\parallel}=R\cdot T_{L-1}\cdot ... \cdot T_1\cdot L
\end{equation}
with
\begin{eqnarray}
  L &=& {\cal L}\otimes\eins\otimes ... \otimes\eins\,,\nonumber\\ 
  R &=& \eins\otimes ... \otimes\eins\otimes{\cal R}\,,\\ 
  T_i &=& \eins\otimes\eins ... \otimes{\cal T}_{i}\otimes\eins ...
  \otimes\eins\,.\nonumber
\end{eqnarray}
The mechanism for stationarity reads now 
\def\edmatrix{\Biggl(\hspace{-1.5mm}
    \begin{array}{c} E \\[-1mm] D \\[-1mm] 0
    \end{array}\hspace{-1.5mm} \Biggr)}
\def\abmatrix{\Biggl(\hspace{-1.5mm}
    \begin{array}{c} \hat{E} \\[-1mm] \hat{D} \\[-1mm] \hat{F}
    \end{array}\hspace{-1.5mm} \Biggr)}
\begin{equation}
\label{paralleldyn}
{\cal T} \, \left[ \abmatrix \otimes \edmatrix \right] \;=\;
\edmatrix \otimes \abmatrix  \,,
\end{equation}
$$
\langle W| {\cal L} \edmatrix \;=\; \langle W| \abmatrix\,,
\hspace{15mm}
{\cal R} \abmatrix |V\rangle\;=\; \edmatrix|V\rangle \,,
$$
with the new third-state matrix  $\hat{F}$. 
This leads to an algebra, which can be found in Appendix \ref{Appalg}.
Note that the last bulk equation $0=D\hat{F}$ excludes a scalar
solution for the algebra. This is consistent with our earlier
observation that there is no simple mean-field solution of the
model.
\\
First, we can check the relations which connect the densities 
at the ends of the chain with the current $J$. This
calculation can be done for arbitrary system sizes and
is presented in Appendix \ref{apprelation}.
\\
Second, it is possible to show that the algebra correctly describes a 
system of three sites. This can be done by taking the expressions for
the weights of the eight possible configurations of the 
stationary state given by (\ref{masterpara}) and by applying   
several times the algebraic rules. Thereby each weight
can be expressed as a 
linear combination of the other weights\footnote{Note that formulas 
for macroscopic variables like the current are quite complicated 
for this small system; the current, for example,  is a 
ratio of polynomials in $\alpha,\beta$ and
$p$ containing 27 additive terms.}.
The resulting system of linear equations turns out to be identical to
those obtained from the transfer matrix $T$. 
\\
Thirdly, it can be checked whether the algebra can be reduced to
a generalized DEHP-algebra of the type
\begin{equation}
\label{haeh}
p\, D\, E=a\, E+b\, D
\end{equation} 
with some numbers $a,b$. This equation induces
certain relations between weights for a system of size $L$ and $L-1$,
 which can be checked using the exact solutions. It turned out that
(\ref{haeh}) cannot be valid.
\\
We found a two-dimensional representation for the bulk algebra \cite{diss}, 
but despite intensive effort, we could not solve the complete 
algebra\footnote{This situation is very similar to \cite{hinsan} where 
a MPA for the ASEP with a defect was proposed.}. Therefore, Monte Carlo
simulations have been performed \cite{dublin}. The results will be
presented in Section \ref{secphasetrans}.

\section{Comparative study of physical quantities}
\label{compasect}

In the following we will investigate the consequences of the mathematical
description developed in the previous section. First we investigate the
ASEP with periodic boundary conditions for the different updates. This
will allow us later to distinguish between ``pure'' bulk effects and 
``boundary-induced'' bulk effects. After that we will derive and compare the
phase diagrams, density profiles and other physical quantities for the
various updates.
\subsection{Periodic boundary conditions}
\label{periodsec}

For random-sequential dynamics the stationary state of the ASEP with
periodic boundary conditions is given by $|P_0\rangle=\frac{1}{Z_L}
{\rm Tr}(A\otimes \dots\otimes A)$  where the elements $E$ and $D$ of 
the vector $A$ satisfy the algebra (\ref{derr1}) and the normalization
is given by $Z_L={\rm Tr\,}C^L$ with $C=E+D$. Since the boundary
equations (\ref{derr2}) and (\ref{derr3}) do not have to be considered, 
it is possible to find a 
one-dimensional representation of the matrices $E$ and $D$ which then 
become real numbers $e$, $d$. The current is calculated from 
$J_L = \frac{p}{Z_L}{\rm Tr}
(C^{j-1}DEC^{L-j-1})=\frac{1}{c}$,
where we have used (\ref{derr1}) and $Z_L=c^L$ with $c=e+d$.
In contrast to the case of open boundary conditions, the density $\rho$ 
is now fixed and the density profile is constant, $\rho = \langle\tau_j
\rangle$, independent of $j$. With 
\begin{equation}
\label{densdef}
\langle\tau_j\rangle=\frac{1}{Z_L} {\rm Tr}(C^{j-1}DC^{L-j})
\end{equation}
one finds $\rho=d/c$. Therefore the current 
is given by
\begin{equation}
\label{Jrsmf}
J(\rho,p) = p\rho(1-\rho)\, .
\end{equation}
This is exactly the mean-field result, because in mean-field approximation
a site is occupied with probability $\rho$ and its right neighbour is
empty with probability $1-\rho$. Since hopping then occurs with probability
$p$, one obtains (\ref{Jrsmf}).
\\
This result is not surprising as the existence of a one-dimensional 
representation implies the absence of correlations between neighbouring 
sites, i.e.\ the MPA reduces to mean-field theory.

We now turn to the ASEP with backward-sequential update on a ring. The state
\begin{equation}
\label{mpg5}
|P_0\rangle \;=\;\frac{1}{Z_L}{\rm Tr}\bigl[A^{\otimes(L-1)}\otimes \hat{A} 
\bigr]
\end{equation}
is obviously translation-invariant and, because of (\ref{Equations}),
stationary (the argument
of the trace is a vector with a product of matrices in each
component; the trace has to be applied to each component).
The algebra can be directly solved with one-dimensional matrices (see
Appendix \ref{appring}). 
The current $J_{\leftarrow}(\rho,p)$ for the ordered-sequential
update is not given by (\ref{current}), but by\footnote{Note that eq.\
(\ref{current_seq}) also applies for the sublattice-parallel update.}
\begin{equation}
\label{current_seq}
J_{\leftarrow}(j)=\frac{p}{Z_L}\langle W|C^{j-1}D\hat{E}C^{L-j-1}|V
\rangle \quad .
\end{equation}
For the case of periodic boundary conditions one finds\footnote{The 
corresponding formula for the ASEP with hopping in both directions can 
be found in App.~\ref{appring}.}
\begin{equation}
\label{strom}
J_{\leftarrow}(\rho,p)=p\rho \frac{1-\rho}{1-p\rho }.
\end{equation}
Fig.\ \ref{X0} illustrates this result. 
Again this result can be obtained directly using a mean-field argument.
The site to the right of an occupied site is empty with probability
$1-(\rho-J_{\leftarrow})$. Here one has to take into account that
the density is reduced by $J_\leftarrow$ after the update of that site.
Therefore the current satisfies $J_{\leftarrow}
=p\rho(1-(\rho-J_{\leftarrow}))$ which leads to (\ref{strom}).
The maximal flow ($p$ fixed) is reached for a density
$\rho^{\rm  max}_\leftarrow(p)=\frac{1}{p}(1-\sqrt{1-p})\geq 1/2$.
The sequential update $T_{\leftarrow}$ ``likes'' high $p$ and high densities.
The particle-hole symmetry can be used to determine these
quantities for $T_{\rightarrow}$ simply by replacing $\rho$ with $1-\rho$:
$T_{\rightarrow}$ ``prefers'' high $p$ and low densities.
\\
It is interesting to compare (\ref{strom}) with the well-known result 
\cite{schreck} for parallel update,
\begin{equation}
J_{\parallel}(\rho,p)=\frac{1}{2}\biggl(1-\sqrt{1-4p \rho (1-\rho)}\biggr).
\end{equation} 
\\ 
For the parallel
update a mean-field theory for the {\em distances} between consecutive
particles becomes exact \cite{COMF}.
Similar results have been obtained recently by Evans \cite{evansneu},
who solved the ASEP with periodic boundary conditions in the presence
of disorder\footnote{Each particle carries its own hopping probability 
$p_j$, see also \cite{krug2,ktitarev}.} with parallel update and a 
particle-sequential update by generalizing the approach of \cite{hakim}. 
\\
$J_{\parallel}(\rho,p)$ is obviously symmetrical with respect to 
$\rho=1/2$. It is maximal at $\rho_{\parallel}^{\rm max}=1/2$ for all 
values of $p$, while $\rho_\leftarrow^{\rm max}$ is always higher than 
$1/2$ (except for $p=0$).
\\
Hence, the maximal currents (for a given $p$) are 
$J_\leftarrow(\rho^{\rm max},p)=\frac{2}{p}(1-\sqrt{1-p})-1$ and
$J_{\parallel}(\rho^{\rm max},p)=\frac{1}{2}(1-\sqrt{1-p})$.
It is intuitively clear that $J_\leftarrow(\rho^{\rm max},p)>
J_{\parallel}(\rho^{\rm max},p)$ holds, and it can be verified easily.

Let us now return to the model with open boundary conditions.

\subsection{Phase diagram}
\label{secphasetrans}

The phase diagram for random-sequential dynamics has been determined
for $p=1$ in \cite{derrida92,derrida93,schdom}. This is no restriction
since, as mentioned before, $p$ only rescales time (and $\alpha \to 
\alpha/p$, $\beta\to\beta/p$). Since it will turn out that the phase
diagrams for the different updates are rather similar we will not
repeat the results for the random-sequential update here. Instead we will
first determine the phase diagram for the ASEP with open boundary conditions 
and discrete time by using the results of Section \ref{seqsect} and discuss
the differences to the random-sequential case later.
In Appendix \ref{appmapp} it is shown how the algebra 
(\ref{BulkAlgebra}),(\ref{BoundaryConditions})
can be projected onto the DEHP-algebra. 
The representations of this algebra (and the resulting phase diagram)
are known for all parameter values $\alpha,\beta,p,q$ 
\cite{sand,essler,mallicksandow}.
Therefore, we have obtained explicit 
expressions for the matrices and vectors, 
and have thus constructed the stationary state of our model; 
we do not have to be concerned about representations any more that
might not satisfy (\ref{trick}) since the projection onto the DEHP-algebra
already covers the whole parameter space.
\\ 
Furthermore, it is 
straightforward to calculate observables such as the current and the
density, at least in principle.
\\
The mapping of App.\ \ref{appmapp} strongly suggests that the well-known 
phase diagram of the ASEP with random-sequential update and stochastic 
hopping in {\it both} directions \cite{sand,essler,mallicksandow} will 
also be valid for 
the ASEP with ordered-sequential update. This is indeed correct and has been
proven directly in \cite{honi}. The phase diagram is shown in
Fig.\ \ref{X1}. 
The mean-field line is the curved dashed line. Also shown are
density profiles calculated from Monte Carlo simulations. 
All well-known features of the phase diagram (high-density phase, low-density
phase, maximum current phase, coexistence line with linear
density profile) are recovered.  
The intersection of the mean-field line (\ref{mean}) with the
line $\alpha=\beta$ defines the endpoint $\alpha_c$ (=$\beta_c$) of
the coexistence line. This yields
\begin{equation}
\alpha_c\, =\, 1-\sqrt{1-p}.
\end{equation}
In the case of deterministic hopping ($p=1$), the
maximum current phase vanishes, and we recover the result of Hinrichsen
\cite{hinri}. 
\\
It is known that for the DEHP-algebra, the mean-field expressions for the
current are exact. Since the mean-field line touches all three phases,
we can calculate the corresponding currents and bulk densities for our
model. Our results are
\begin{equation}
J(\alpha,\beta,p)= \begin{cases} \frac{\alpha}{p}\frac{p-\alpha}
{1-\alpha} & \text{in the low-density phase}, \\ \frac{\beta}{p}
\frac{p-\beta}{1-\beta} & \text{in the high-density
phase}, \\ \frac{1-\sqrt{1-p}}{1+\sqrt{1-p}} & \text{in the maximum current 
phase}, \end{cases}
\end{equation}
which is in excellent agreement with our numerical data.\\
Since the relation $J(\alpha,\beta,p)=\beta\rho(\alpha,\beta,p,x=L)$
is exact, we immediately get the bulk density in the
high-density phase: $\rho_{\rm bulk}=\frac{1}{p}\frac{p-\beta}{1-\beta}$.
The bulk density in the low-density phase will be obtained below.
Fig.\ \ref{X3} shows a space-time diagram for a point on the coexistence line
($p=0.75,\alpha=\beta=0.3$) produced by the Monte Carlo
simulation. The well-known fluctuating shock can nicely be observed.
\\
Typical space-time diagrams for the different phases can be found
in Fig.\ \ref{X4}. The ``jams'' in the high-density phase move from the
right to the left. In the low-density phase, groups of particles
(small jams) move from the left to the right.

For the case of parallel dynamics extensive Monte Carlo 
simulations have been performed \cite{dublin} which revealed that the
phase diagram looks essentially the same as before (Fig.\ \ref{X1}). 
Again, we find a high- (low-) density and a maximum current phase and a linear
density profile along the line $\alpha=\beta$, until the 2-cluster
line (\ref{meanfield}) is touched.
\\
Since we have a particle-hole symmetry in the model, the density for
$\alpha=\beta$ and odd lattice sizes has to be $1/2$ for 
site $\frac{L+1}{2}$. Therefore, the 'bulk density' 
$\rho(\alpha,\beta,p,x=\frac{L+1}{2})$ in
the maximum current phase must be $1/2$.
\\
Equations (\ref{cluster}) and (\ref{meanfield}) can be used to calculate
the current\footnote{For the parallel update the current is given by
(\ref{current}).}
\begin{equation}
J(\alpha,\beta,p)= \begin{cases} {\alpha}\frac{p-\alpha}{p-{\alpha}^2}
  & \text{in the low-density phase}, 
\\ 
\beta\frac{p-\beta}{p-{\beta}^2} & \text{in the high-density phase},
  \\ 
\frac{1-\sqrt{1-p}}{2} & \text{in the maximum current phase}, \end{cases}
\end{equation}
and the remaining bulk densities $\rho_{\rm bulk}=\frac{\alpha
  (1-\alpha)}{p-{\alpha}^2}$ (low-density) and $\rho_{\rm
  bulk}=\frac{p-\beta}{p-{\beta}^2}$ (high-density phase). These results
are exact on the 2-cluster line (\ref{meanfield}). For general values of
$\alpha$ and $\beta$ we find an excellent agreement with the numerical data. 
Fig.\ \ref{X4} presents typical space-time diagrams in the various phases. The 
particle-hole attraction is apparent.

Note that for every update the maximal current $J(\rho^{\rm max},p)$ 
on the ring (as calculated in Section \ref{periodsec})
is equal to the value of the current in the maximum current phase for 
that update. The value of $\rho^{\rm max}$ gives the corresponding 
bulk density.
This shows that the current in the maximum current phase is determined
by the 'capacity' of the ring. 


\subsection{Density profiles and phase transitions}

For the parallel update we performed extensive Monte Carlo simulations 
in order to determine quantities like density profiles and correlation 
functions. In this section we will compare these results with
those obtained analytically and numerically for the other update types.

The results presented in Section \ref{secphasetrans} show that one can 
distinguish at least three
different phases with respect to the flow. In the low- (high-) density
phase the current depends 
for a given value of $p$ only on $\alpha$ ($\beta$) and in the maximum
current phase the current is independent of the chosen in- and output
rates. This basic structure of the phase diagram is the same for all
types of update although the bulk properties change drastically.
 
A more detailed analysis of ASEP with random-sequential update has shown 
\cite{schdom} that the system is governed by two independent length scales
$\xi_\alpha$ and $\xi_\beta$, which represent the influence of
the boundaries. At the critical value of $\alpha$ and $\beta$ these length
scales diverge and a phase transition occurs. This divergence produces
two additional phases compared to the mean-field results. Thus the
parameter space is divided into five different phases. The low-density
phases AI and AII, where the bulk properties are determined by the
value of $\alpha$, the high-density phases BI and BII, where the output
rate $\beta$ determines the flow, and the maximum current phase where
the flow is independent of $\alpha$ and $\beta$. The phases AI and AII
(BI and BII) are distinguished by the behaviour of the density profile
near the boundaries (see below).

We checked this scenario for the parallel and ordered-sequential
update. In order to avoid difficulties due to the $p$-dependent scale
factor of the phase diagram we analyzed the density profiles for
$p=0.75$. 
Our simulation results are obtained for systems with 
$320$ sites. Since finite-size corrections are rather small,
this size is already sufficient to obtain the behaviour 
in the limit of large $L$.  

We first measured the density profile in the low-density phases AI and AII. 
In \cite{schdom,derrida93} it has been shown that  for large $L$ the 
asymptotic behaviour near the boundaries changes from a pure exponential 
decay in the phase AI to the enhanced exponential decay according to
$\exp(-(L-x)/\xi_\alpha)/\sqrt{L-x}$ in the phase AII. This change of the
behaviour near the boundaries is due to the divergence of the length scale
$\xi_\beta$ at the transition line. $\xi_\beta$ remains infinite throughout
the whole phase AII. Unfortunately, the divergence of
$\xi_\beta$  cannot be calculated directly because the relevant length
scale $\xi$, which determines the exponential decay in AI, is given by 
\begin{equation}
\xi^{-1} = \bigl|\xi_{\alpha}^{-1} - \xi_{\beta}^{-1}\bigr|.
\label{corr_length}
\end{equation}
Thus the correlation length $\xi$ remains finite at the transition line from
AI to AII. In
contrast to the divergence of $\xi_\beta$, the asymptotics of the density
profile can be checked against the numerical results.
Fig.~\ref{a12prof} shows that the density profile for $\alpha= 0.40 $, 
$\beta= 0.42$ and parallel update can be
nicely fitted by a pure exponential decay, but within the phase AII
the enhanced exponential form has to be used (see also the insert in
Fig.\ \ref{X1}). Another characteristic line crossing the low-density phase
AII is the mean-field line, where the density profile is completely flat.
This line separates the monotonously decreasing density profiles from
monotonously increasing profiles, but the asymptotic behaviour is left
unchanged. The behaviour of the density profiles in the high-density
phases BI and BII can be obtained using the particle symmetry of the
model.  

The transition between the high- and low-density phases is driven by the
diverging length scale $\xi$. This divergence occurs although both
length scales are finite, because at the transition line $\xi_\alpha$ and 
$\xi_\beta$ coincide. The qualitative agreement of the density profile
strongly suggest that the relevant length scale in the low-density
phase is given by (\ref{corr_length}) also for discrete-time updates.
Moreover, the comparison of the numerically estimated correlation
length shows that the correlation lengths of the parallel and 
ordered-sequential update are identical and differ from those obtained 
for the random-sequential only by a constant factor (Fig.\ \ref{a1length}). 
Exactly at the transition line $\alpha = \beta $, one finds a linear
density profile as shown in
Fig.\ \ref{a1b1trans}. This linear profile is result of a fluctuating
shock front which separates for a given time a high-density region from
a low-density region. The position of the shock front fluctuates
through the whole lattice such that a time average over all single
time profiles gives a linear profile.

The transition from the low- (high-) density phase AII (BII) to the
maximum current phase is characterized by a change of the asymptotic
behaviour of the density profile from the enhanced exponential
to a pure algebraic decay. The transition is driven by the divergence of 
the length scale $\xi_\alpha$ ($\xi_\beta$) for $\alpha\rightarrow \alpha_c$
($\beta\rightarrow\beta_c$). Both length scales are infinite
not only at the transition line but for all values of  $\alpha$ and
$\beta$ larger than the critical value. Therefore an algebraic
decay of the density profile can be observed in the whole maximum
current phase. 
Fig.~$\ref{a2length}$ shows the diverging correlation length  for different
update types.    Again the length scales of the updates with discrete
time  agree and the length scales produced by the random-sequential  
update are larger if one considers the same in- and output rates.

The flow is generated by the (10)-clusters (the mobile pairs), while the 
other 2-cluster configurations exclude hopping of particles. Therefore, 
we measured density profiles of the probabilities $P_{\tau_i\tau_{i+1}}$
of 2-cluster configurations $\tau_i\tau_{i+1}$. 
One gets a flat profile of the mobile pairs $P_{1_i0_{i+1}}$ 
(see Fig.~\ref{pairs}) which is a consequence of eq.~(\ref{current}) 
for the current in the case of parallel dynamics\footnote{This is not true
for the ordered-sequential update, since in this case the current depends
on the local defect generated by the update. Hence, the current 
depends on $D\hat{E}$ (see (\ref{current_seq})) instead of $DE$ (see 
(\ref{current})).}. The nontrivial part of the density profile is produced 
by the immobile pairs. The identity $P_{1_i0_{i+1}}= P_{0_i1_{i+1}}$, 
which is already known from the periodic system \cite{traffic}, 
is only true for flat density profiles. 
One observes qualitatively the same behaviour for the
random-sequential, but not for the ordered-sequential update.
\\
For the ordered-sequential update the transport of a local defect
changes the behaviour of the $P_{\tau_i\tau_{i+1}}$ profiles: 
The density of particles  in front of hole at the defect site (in the
case of backward-sequential update) determines the flow and therefore
none of the 2-cluster configurations is in general translation-invariant  
after a sweep through the lattice.


\section{Discussion}
\label{discsect}

We have presented an extensive comparative study of the ASEP with
different updates. The purpose of this investigation is threefold:
First of all, despite the importance of the ASEP, not much has been known
about its properties for discrete-time updates.
Second, we tried to obtain a better understanding of the similarities
and differences of the different updates. Finally, there is also a 
practical aspect. In \cite{ito,lecaer} it has been shown 
that different dynamics perform quite differently in Monte Carlo 
simulations. In order to save computational time it might therefore
be useful to choose a certain update. Then it is also necessary to
know how to translate the results into those for other updating
schemes.

The main tool that we used in our investigations was the MPA which allows
for an analytical solution for the cases of random-sequential, 
ordered-sequential and sublattice-parallel updates. We also proposed
a MPA for the important case of parallel dynamics, but unfortunately
we were not able to find a general representation of the resulting
matrix algebra. Therefore extensive Monte Carlo simulations have been
performed in order to determine the phase diagram. These numerical
results, together with an analytical solution for a special line in
parameter space, allowed us to conjecture analytical expressions for the
phase diagram.

Our results show that the phase diagram has the same basic structure for all 
the updates investigated here. One finds three different phases characterized 
by the value of the current. For $\alpha > \beta$ and $\beta < \beta_c(p)$
the system is in the so-called high-density phase. Here the current
depends only on the removal rate $\beta$ since particles are inserted
much more efficiently than they are removed. Just the opposite situation
is found in the low-density phase, $\alpha < \beta$ and $\alpha < 
\alpha_c(p)$. Here the removal is much more effective than the
insertion and the current depends only on $\alpha$. Note that for each
update $\alpha_c(p)=\beta_c(p)$ and the functional forms of the currents
in the high- and low-density phases are identical (see Table 1).

Finally, for $\alpha > \alpha_c(p)$ and $\beta > \beta_c(p)$ one finds the
maximum current phase. Here the current is independent of $\alpha$ and
$\beta$. Both insertion and removal are so effective that the current is
only limited by the ``bulk capacity''. Indeed, the current in this phase
is identical to the maximal current of the corresponding system with
periodic boundary conditions.

Phase transitions between these phases are driven by diverging length
scales $\xi_\alpha$ or $\xi_\beta$ within the high- and low-density phases.
These length scales depend on the rates $\alpha$ and $\beta$, respectively.
In contrast to that, the periodic systems exhibit only extremely short-ranged
correlations. The strongest correlations are found for the parallel update,
but even here already the 2-cluster approximation is exact. This leads to
exponentially decaying correlation functions with a rather short correlation
length (except for $\rho\approx 1/2$ and $p\approx 1$) \cite{eisi}.
So the long-ranged
correlations found for the open system are due to the boundary conditions
and the transitions are genuine ``boundary-induced phase transitions''.
This makes it also understandable why the phase diagrams for the different
updates look so similar, the only difference being the location of the
transitions and the functional form of the observables.
Furthermore the numerical analysis shows that the update type does not
change the ``universal'' properties of the model:  We observe 
the same asymptotics of the density profiles in the different
phases and also the qualitative behaviour of the length scales near the
phase boundaries does not depend on the update. For the discrete
updates the length scales agree even quantitatively. 

Recently, for the case of periodic boundary conditions there has been some
progress in the understanding of the differences between random-sequential
and parallel update \cite{eden}. For the latter so-called ``Garden of Eden''
(GoE) states\footnote{Here on has to distinguish between particles which 
moved in the previous timestep (velocity $1$) and particle that did not move
(velocity $0$).} exist. These states can not be reached dynamically, they 
do not have a predecessor. By eliminating these states one finds in the reduced
configuration space that mean-field theory becomes exact, as it is for the 
case of random-sequential update (but here in the full configuration space 
since Garden of Eden states do not exist). Therefore the existence of 
GoE states is the reason for the correlations in the ASEP 
with parallel dynamics and periodic boundary conditions. Since the bulk
dynamics for the ASEP with open boundaries is the same as for the
periodic case, we expect the GoE states to play an important role also
in that case.

In this paper, valuable information has been gained by means
of low-dimensional representations of the matrix algebras.
A one-dimensional  representation 
clearly corresponds to a simple mean-field approximation.
However, in the case of two (or higher) dimensional representations,
nothing is known about the {\it physical} interpretation of
these solutions. For example, it could be possible that
there is a close connection
between cluster approximations and these representations; we
remark that it is possible to write the expectation values of 
densities and correlations of any exact (stationary) solution of a 
2-cluster approximation as a product of two-dimensional matrices
precisely  in the form resulting from a MPA.

Recent investigations have shown that the ASEP is capable of reproducing
the essential features of traffic in a city such as Geneva
{\cite{chopard}}. The authors studied a simple extension of the
one-dimensional ASEP to two dimensions\footnote{In \cite{duis,dallas1,dallas2}
other generalizations of the ASEP have been used for simulations of urban 
traffic in Duisburg and Dallas.}. Therefore, it would be most
interesting to generalize the MPA to higher dimensions\footnote{See
\cite{tadaki} for a numerical investigation of a two-dimensional ASEP.}. 
A first step
would be to study the ASEP on a ladder. The analytical method used in
this paper is directly applicable to ``stochastic'' ladders
\cite{rajrau}.  Since the groundstates of certain quantum spin ladders
have been constructed recently \cite{kolezhuk,zitt2} using optimum 
groundstates (see Section \ref{sectoptimum}),
there seems to be a good chance to find low-dimensional representations of
the corresponding algebras. This would lead to a better analytical
understanding of the fascinating phenomena that occur in higher-dimensional
nonequilibrium systems \cite{zia,ziabook}.\\[0.5cm]

{\Large {\bf Acknowledgments}}\\[0.3cm]
This work has been performed within the research program of the 
Sonderforschungsbereich 341 (K\"oln-Aachen-J\"ulich).
We would like to thank A.\ Honecker, A.\ Kl\"umper, H.\ Niggemann,
I.\ Peschel and G.\ Sch\"utz for useful discussions and the
HLRZ at the Forschungszentrum J\"ulich for generous allocation
of computing time on the Intel Paragon XP/S~10..

\begin{appendix}
\renewcommand{\theequation}{\Alph{section}.\arabic{equation}}
\section{Mapping of the ordered-sequential algebra onto the DEHP-algebra}
\label{appmapp}
\setcounter{equation}{0}
We here treat the general case where particles are also allowed to 
hop to empty sites on the left with probability $q$.
In this case the bulk algebra (\ref{BulkAlgebra})
is generalized to 
\begin{eqnarray}
\label{BulkAlgA}
[E,\hat{E}] =[D,\hat{D}] &=& 0 ,\nonumber \\
(1-q)E\hat{D}+pD\hat{E} &=& \hat{E}D ,\\
qE\hat{D}+(1-p)D\hat{E} &=& \hat{D}E ,\nonumber
\end{eqnarray}
with the boundary conditions (\ref{BoundaryConditions}).
We first note that
\begin{equation}
\label{ccommut}
[E+D,\hat{E}+\hat{D}]=0
\end{equation}
holds for all values of $p,q$.
\\ 
The crucial step is now to demand
\begin{eqnarray}
\label{trick}
\hat{E}&=& E+\lambda \eins \,,\nonumber\\
\hat{D} &=&D-\lambda \eins 
\end{eqnarray}
with some (real) number $\lambda$.
Note that this is the simplest way to satisfy (\ref{ccommut}).
\\
Now, we must show that a solution (representation) of
the algebra (\ref{BulkAlgebra}), (\ref{BoundaryConditions}) plus
equations (\ref{trick}) exists for all possible values of the parameters
$\alpha,\beta,p,q$.
\\ 
(\ref{trick}) reduces the algebra from eight equations 
to three equations:
\begin{eqnarray}
pDE-qED &=& \lambda(1-q)E+\lambda(1-p)D\nonumber\,,\\
\alpha\langle W|E &=& \lambda(1-\alpha)\langle W|\,,\label{paralsub}\\
\beta D|V \rangle &=& \lambda |V \rangle \nonumber.
\end{eqnarray}
We define
\begin{eqnarray}
\widetilde{D}&:=&\lambda(1-p)D \,,\nonumber\\
\widetilde{E}&:=&\lambda(1-q)E \,, \label{edtilde}\\
\lambda^2&:=&\frac{1}{(1-q)(1-p)}\,, \nonumber
\end{eqnarray}
and rewrite (\ref{paralsub}) as
\begin{eqnarray}
  p\widetilde{D}\widetilde{E}-q\widetilde{E}\widetilde{D} &=&
  \widetilde{E}+ \widetilde{D}\,, \nonumber\\ \alpha (1-p)\langle W|
  \widetilde{E}&=& (1-\alpha)\langle W|\,,\label{fullalgebra}\\ 
  \beta(1-q)\widetilde{D}|V \rangle &=& |V \rangle.\nonumber
\end{eqnarray}
This is the algebra for the ASEP with random-sequential update and hopping
in both directions (with probability $p$ and $q$, respectively), but
with the same local transfer matrix and the same boundary conditions 
as in our model. 
The algebra was solved for $p=1$ and $q=0$ by Derrida et al.\
\cite{derrida93} with infinite-dimensional matrices. Note that the
vectors $\langle W|$ and $|V\rangle$ of their solution have to be
rescaled with $\frac{1-\alpha}{1-p}$ and $\frac{1}{1-q}$, respectively. 
A thorough discussion of the algebra (\ref{fullalgebra}) can be found 
in \cite{sand,essler,mallicksandow,sasa}.

We write down an explicit representation of the algebra (\ref{paralsub})
with $\lambda=1$ in the case $q=0$:
\begin{equation}
{D} \;=\; \frac{1}{p}\left(
\begin{array}{cccccc}
\frac{p}{\beta} & a_1 & 0 & 0 & \cdot \\
0 & 1 & 1 & 0 & \cdot \\
0 & 0 & 1 & 1 & \cdot \\
0 & 0 & 0 & 1 & \cdot \\
\cdot & \cdot & \cdot & \cdot & \cdot \\
\end{array} \right)\,, 
\end{equation}\begin{equation}
{E} \;=\; \frac{1}{p}\left(
\begin{array}{cccccc}
\frac{p(1-\alpha)}{\alpha} & 0 & 0 & 0 & \cdot \\
a_2 & 1-p & 0 & 0 & \cdot \\
0 & 1-p & 1-p & 0 & \cdot \\
0 & 0 & 1-p & 1-p & \cdot \\
\cdot & \cdot & \cdot & \cdot & \cdot \\
\end{array} 
\right)\,,\end{equation}\begin{equation}
\langle W|=(1,0,0,0,\ldots),\quad |V\rangle=\left( \hspace{-1.5mm} 
\begin{array}{c} 1 \\ 0\\ 0\\ 0\\ \vdots\\ \end{array} 
\hspace{-1.5mm}  \right)\quad ,
\end{equation} 
\begin{equation}
a_1a_2=\frac{p}{\alpha\beta}[(1-p)-(1-\alpha)(1-\beta)]\,.
\end{equation}
As expected, constraint (\ref{mean}) leads to an effectively one-dimensional
representation.

\section{Symmetries of the density profiles for the ordered-sequential updates}
\label{appsymm}
\setcounter{equation}{0}
As shown in Section \ref{secnonc} the stationary state for 
$T_{\rightarrow}$ is simply given by (\ref{statright}).
When inserting (\ref{edtilde}) into (\ref{statright}), we get
a connection between the density profiles\footnote{In the following
we will denote the local density $\langle\tau_x\rangle$ at site $x$ by 
$\rho(\alpha,\beta,p,x)$ in order to stress the dependence on the
other parameters.}
of the stationary states produced by $T_{\leftarrow}$ and $T_{\rightarrow}$:
\begin{equation}
\rho_{\rightarrow}(\alpha,\beta,p,x)=\rho_{\leftarrow}(\alpha,\beta,p,x)-
\lambda\frac{Z_{L-1}}{Z_L} \, .
\end{equation}
One also has  $J(\alpha,\beta,p)=\beta\rho(\alpha,\beta,x=L)$ 
which leads to
\begin{equation}
\label{rholr}
\rho_{\rightarrow}(\alpha,\beta,p,x)=\rho_{\leftarrow}
(\alpha,\beta,p,x)-J(\alpha,\beta,p)\, .
\end{equation}
$\rho_{\rightarrow}(\alpha,\beta,p,x)$ is therefore always
lower than $\rho_{\leftarrow}(\alpha,\beta,p,x)$. 
The current $J$ is not $x$-dependent. This
means that the density profile of the ordered-sequential model for a 
given set of parameters $\alpha,\beta,p$ is, up
to a constant (the current), the same for both directions of 
the order of update. The stationary states produced by
$T_{\rightarrow}$ and $T_{\leftarrow}$ for a given set of parameters
{\it are always in the same phase}.
It is obvious that the ``crucial step'' (\ref{trick})
is {\it the simplest} way to obtain such a behaviour. \\
When using the particle-hole symmetry
\begin{equation} 
\label{phsymmet}
 \rho_{\rightarrow}(\alpha,\beta,p,x)=1-\rho_{\leftarrow}(\beta,
\alpha,p,L-x+1)\, ,
\end{equation} 
we immediately see that the density profile on the line $\alpha=\beta$ 
has to be symmetric with respect to point $L/2$. For this 
case, we further obtain
$\rho_{\leftarrow}(\alpha,p,\frac{L+1}{2})=\frac{1+J(\alpha,p)}{2}$.
Finally, by using (\ref{rholr}) and (\ref{phsymmet}), and the results for the 
current and $\rho_{\rm bulk}$ in the high-density region, the bulk 
density in the low-density region $\rho_{\rm bulk}=\alpha /p$ is obtained.

\section{Symmetric diffusion}
\label{appsymmdiff}
\setcounter{equation}{0}

We briefly discuss the case of symmetric diffusion ($p=q$) for
the ordered-sequential update.
The density $\rho(\alpha,\beta,p,x)$ at site $x$ ($1\le x\le L$)
is given by
\begin{equation}
\label{klar} 
\rho(\alpha,\beta,p,x)=\frac{1}{Z_L}\langle W| C^{x-1}DC^{L-x}|V\rangle.
\end{equation}
(\ref{paralsub}) and (\ref{edtilde}) yield
\begin{equation}
\label{bloede}
[D,C]=[D,E+D]=\frac{1}{p}(E+D)=\frac{1}{p}C.
\end{equation}
This means that the density can be calculated immediately by
commuting $D$ with the help of (\ref{bloede}) through the chain in
(\ref{klar}).
By using the boundary conditions, the density can easily be calculated, 
which in turn makes it possible to estimate the current \cite{honi}.
It is intuitively clear that 
the current vanishes in the thermodynamical limit $L\rightarrow
\infty$ for arbitrary values of $\alpha,\beta,p$. 
\section{ASEP with ordered-sequential update on a ring}
\label{appring}
\setcounter{equation}{0}
Since we expect that the periodic system can be described by an
one-dimensional representation of the algebra (\ref{BulkAlgA}), we
are looking for solutions of (\ref{BulkAlgA}) with real numbers 
$e,d,{\hat e},{\hat d}$:
\begin{eqnarray}
(1-q)e{\hat d}+pd{\hat e} & = & {\hat e}d \, ,\label{oooo}\\
qe{\hat d}+(1-p)d{\hat e}& = & {\hat d}e \, .
\end{eqnarray}
The normalized density $\rho$ is given by
\begin{equation}
\rho=\frac{d(e+d)^{L-2}({\hat e}+{\hat d})}{(e+d)^{L-1}({\hat e}+{\hat
    d})} \,=d\frac{1}{(e+d)} .
\end{equation}
There is some freedom how to choose $e,d,{\hat e},{\hat d}$. We
set $e+d={\hat e}+{\hat d}=1$. Therefore, (\ref{oooo}) yields 
\begin{equation}
{\hat e}=\frac{\rho(q-1)+1-q}{\rho(q-p)+1-q} \, .
\end{equation}
The current $J$ is given by $J=pd{\hat e}-qe{\hat d}$ and one obtains
\begin{equation}
J(p,q)=\rho\frac{(\rho-1)(q-p)}{\rho(q-p)+1-q}\, .
\end{equation}
Obviously, $J(p,q)=0$ for the case of symmetric diffusion, $p=q$.

Finally, we like to point out that the fact that the periodic system 
is described by a one-dimensional representation implies that in this
case mean-field theory is exact.

\section{$T_{\parallel}$ in a three-state notation}
\label{app3state}

\setcounter{equation}{0}
The operators ${\cal R}$,${\cal L}$ can be written as: 
\begin{equation}
  {\cal R} \;=\; \left(
\begin{array}{ccc}
  1 & \beta & 0 \\ 
  0 & 1-\beta & 1 \\
  0 & 0 & 0
\end{array} \right) \,,
\hspace{10mm}
{\cal L} \;=\; \left(
\begin{array}{ccc}
  1-\alpha & 0 & 0 \\ 
  0 & 1 & 0 \\
  \alpha & 0 & 0
\end{array} \right)\,.
\end{equation}
The basis chosen for ${\cal R}$ and ${\cal L}$ is $(0,1,-1)$.\\ 
The local transfer-matrix ${\cal T}_{i}$
for any pair of sites $(i,i+1)$ is nine-dimensional. Since the third 
state ($-1$) may not appear {\it after} the update of the whole chain, 
the last four rows and every third column are irrelevant
and here set to zero:
\begin{equation}
\label{Interaction2}
{\cal T}_{i} \;=\; \left(
\begin{array}{ccccccccc}
 1 & 0 & 0 & 0 & 0 & 0 & 0 & 0 & 0 \\
 0 & 1 & 0 & 0 & 0 & 0 & 0 & 0 & 0 \\
 0 & 0 & 0 & p & 0 & 0 & 0 & 0 & 0 \\
 0 & 0 & 0 & 1-p & 0 & 0 & 1 & 0 & 0 \\
 0 & 0 & 0 & 0 & 1 & 0 & 0 & 1 & 0 \\
 0 & 0 & 0 & 0 & 0 & 0 & 0 & 0 & 0 \\
 0 & 0 & 0 & 0 & 0 & 0 & 0 & 0 & 0 \\
 0 & 0 & 0 & 0 & 0 & 0 & 0 & 0 & 0 \\
 0 & 0 & 0 & 0 & 0 & 0 & 0 & 0 & 0 
\end{array} \right)\,.
\end{equation}
The basis is $(00,01,0-1,10,11,1-1,-10,-11,-1-1)$. 

\section{Matrix algebra for parallel dynamics}
\label{Appalg}
\setcounter{equation}{0}

The MPA for parallel update  proposed in Section \ref{parsect}
leads to the following algebra:
\begin{eqnarray}
\hat{E}E & = & E\hat{E}, \label{a1}\\
\hat{E}D & = & E\hat{D}, \label{a2}\\
p\hat{D}E &=& E\hat{F}, \label{a3}\\
(1-p)\hat{D}E + \hat{F}E &=& D\hat{E},\label{a4}\\
\hat{D}D+\hat{F}D &=& D\hat{D}, \label{a5}\\
0 &=& D\hat{F},\label{a6}
\end{eqnarray}
and the boundary conditions
\begin{eqnarray}
   \langle W| E (1-\alpha) \;=\; \langle W| \hat{E}\quad ,  \label{b1}\\
   \langle W| D \;=\; \langle W| \hat{D} \quad, \label{b2}\\
   \langle W| \alpha E \;=\; \langle W| \hat{F} \quad, \label{b3}\\
   \hat{E}+\beta\hat{D}|V\rangle \;=\; E|V\rangle\quad , \label{b4}\\
   (1-\beta)\hat{D}+\hat{F}|V\rangle \;=\; D|V\rangle \quad .\label{b5}
\end{eqnarray}
Note that the last bulk equation $0=D\hat{F}$ excludes a scalar
solution for the algebra. This is consistent with our earlier
observation that there is no simple mean-field solution of the
model.

\section{Check of current-density relations}
\label{apprelation}
\setcounter{equation}{0}

Defining $C=E+D$ and $\hat{C}=\hat{E}+\hat{D}+\hat{F}$ it is easy to
see that
\begin{equation}\label{commu}
[C,\hat{C}]\;=\;0
\end{equation}
holds. This is implied by probability conservation (the columns of
${\cal{T}}_i$ add up to one) and the exchange mechanism (\ref{paralleldyn}).
The boundary equations lead to
\begin{equation}\label{wech}
\langle W| C \;=\; \langle W|\hat{C}
\end{equation}
and
\begin{equation}
C|V\rangle \;=\; \hat{C}|V\rangle\quad .
\end{equation}
The bulk equations (\ref{a2}), (\ref{a5}) give
\begin{equation}
\label{prac}
C\hat{D}=E\hat{D}+D\hat{D}=\hat{D}D+\hat{F}D+\hat{E}D=\hat{C}D\quad.
\end{equation}
We can now check the relations which connect the densities 
 at the ends of the chain and the current $J$ 
which has to be constant throughout the chain:
\begin{eqnarray}
J(\alpha,\beta,p) & = &\beta\rho(\alpha,\beta,p,x=L)\, \\
J(\alpha,\beta,p) & = &\alpha[1-\rho(\alpha,\beta,p,x=1)]\, .
\end{eqnarray}
Note that the first (second) equation would not hold for the sequential update
from the left (right) to the right (left) and that in fact we have
only to prove one of these equations because we can make use of the 
particle-hole symmetry of the model. We therefore write, using the
algebra (equations (\ref{b5}), (\ref{a3})):
\begin{eqnarray}
\rho(\alpha,\beta,p,x=L) & = & \frac{1}{Z_L} \langle
W|C^{L-1}D|V\rangle \\
 & = & \frac{1}{Z_L} (1-\beta)\langle W|C^{L-1}\hat{D}|V\rangle\;+\frac{1}{Z_L}
 p\langle W|C^{L-2}\hat{D}E|V\rangle. \nonumber
\end{eqnarray}
Making use of (\ref{prac}), commuting $\hat{C}$ to the left end of the
chain (\ref{commu}), transforming it to ${C}$ there (\ref{wech}), and
using $ J(\alpha,\beta,p)=\frac{1}{Z_L} p\langle W|C^{L-2}DE|V\rangle$ we get 
\begin{eqnarray}
\rho(\alpha,\beta,p,x=L) & = & \frac{1}{Z_L} (1-\beta)\langle
W|C^{L-1}D|V\rangle\;+\frac{1}{Z_L} p\langle W|C^{L-2}DE|V\rangle
\nonumber \\
 & = & (1-\beta)\rho(\alpha,\beta,p,x=L)+ J(\alpha,\beta,p),
\end{eqnarray}
which is the desired result.

\end{appendix}

\newpage
\begin{center}
{\large Table}
\end{center}
\ \\[1cm]
\begin{tabular}{cccc}
 & rand.-sequential & ordered-seq. ($T_{\leftarrow}$) & parallel \\
 \hline
 & & & \\
low-density & $J=p\alpha(1-\alpha)$ &
 $J=\frac{\alpha}{p}\frac{p-\alpha}{1-\alpha}$ &
 $J=\alpha\frac{p-\alpha}{p-{\alpha}^2}$ \\
 phase & $\rho=\alpha$ & $\rho=\frac{\alpha}{p}$ &
 $\rho=\frac{\alpha(1-\alpha)}{p-{\alpha^2}}$ \\
& & & \\ \hline & & & \\
high-density & $J=p\beta(1-\beta)$ &
 $J=\frac{\beta}{p}\frac{p-\beta}{1-\beta}$ &
 $J=\beta\frac{p-\beta}{p-{\beta}^2}$ \\
phase & $\rho=1-\beta$ & $\rho=\frac{1}{p}\frac{p-\beta}{1-\beta}$ &
 $\rho=\frac{p-\beta}{p-{\beta}^{2}}$ \\ & & & \\ \hline & & & \\
max. current & $J=\frac{p}{4}$ & $J=\frac{1-\sqrt{1-p}}{1+\sqrt{1-p}}$
 & $J=\frac{1-\sqrt{1-p}}{2}$ \\
phase & $\rho=\frac{1}{2}$ &
 $\rho_{L/2}=\frac{1}{1+\sqrt{1-p}}$ &
 $\rho_{L/2}=\frac{1}{2}$ \\ & & & \\ \hline \hline
 & & & \\
critical rate  &  $\alpha_c =\frac{p}{2}$  & $\alpha_c
=1-\sqrt{1-p}$ &  $\alpha_c =1-\sqrt{1-p}$ \\ 
 & & & \\ 
\hline
\end{tabular}
\\ 
\phantom{as}\\[0.7cm]
Table 1: Comparison of currents and (bulk) densities in the three phases
for the different updates. The bulk densities for $T_{\rightarrow}$ are 
given by $\rho-J$ for $T_{\leftarrow}$. The currents for $T_{\rightarrow}$ 
and $T_{\leftarrow}$ are identical.

\newpage
\begin{center}
{\large Figure Captions}
\end{center}
\begin{description}
\item[Fig.\ 1] Definition of the ASEP.
\item[Fig.\ 2] (i) Ordered-sequential update from the right to the left,
(ii) from the left to the right, (iii) and the sublattice-parallel update.
\item[Fig.\ 3] Fundamental diagram (flow vs.\ density) for the ASEP with 
ordered-sequential update and periodic boundary conditions. The vertical
lines indicate the location of $\rho^{\rm  max}_\leftarrow(p)$.
\item[Fig.\ 4] Phase diagram for the ASEP with ordered-sequential update 
$T_{\leftarrow}$ for $p=0.5$. The mean-field line (\ref{mean}) is curved 
dashed line. The straight dashed lines are the boundaries between the
phase AI and AII (BI and BII). The inserts show density profiles obtained 
from Monte Carlo simulations.
\item[Fig.\ 5] The space-time diagram for
  $p=0.75,\alpha=\beta=0.3$. The microscopic shock moves freely on the
  chain.
\item[Fig.\ 6] Typical space-time diagrams for the various phases.
\item[Fig.\ 7] Density profiles for parallel update (p=0.75) in the 
      low-density phases AI and AII. The insert compares the  exact
      asymptotic form with a pure exponential decay.
\item[Fig.\ 8]Log-Log plot of the correlation length for
      different types of updates (OS=ordered-sequential, PARA=parallel,
      RS=random-sequential). The length scales are obtained from
      an exponential fit of the density profile, $\alpha$  and $\beta$
      are chosen such that $\alpha + \beta =0.6$ holds
      $(\alpha_c=0.3)$. 
\item[Fig.\ 9] Density profiles near the first order transition
      at $\alpha  = \beta = 0.3$ for the random sequential
      update using the exact results of \cite{derrida93}. Again
      $\alpha + \beta =0.6$ holds.
\item[Fig.\ 10] Divergence of the correlation length $\xi_\alpha$
      near the transition to the maximum current phase ($\beta =
 0.75$).
\item[Fig.\ 11] Density profile of the pair probabilities
      $P_{\tau_i\tau_{i+1}}$ for the parallel update in the low
      density phase AII ($\alpha = 0.40, \beta = 0.75, p = 0.75$).
\end{description}
\newpage
\begin{figure}[ht]
\label{asepfig}
\centerline{\psfig{figure=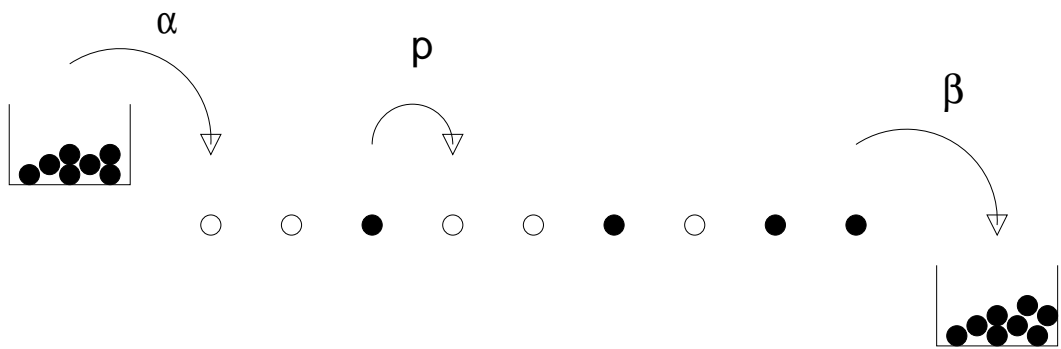,bbllx=60pt,bblly=220pt,bburx=550pt,bbury=410pt,
height=6cm}}
\caption{Definition of the ASEP.}
\end{figure}
\begin{figure}[ht]
\label{updates}
\centerline{\psfig{figure=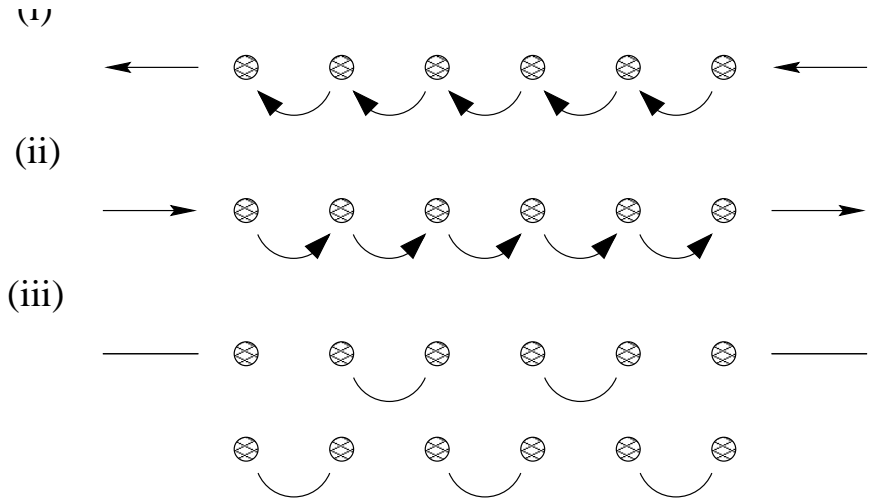,bbllx=90pt,bblly=300pt,bburx=460pt,bbury=650pt, height=11cm}}
\caption{(i) Ordered-sequential update from the right to the left,
(ii) from the left to the right, (iii) and the sublattice-parallel update.}
\end{figure}
\vfill\eject
\begin{figure}[ht]
\centerline{\psfig{figure=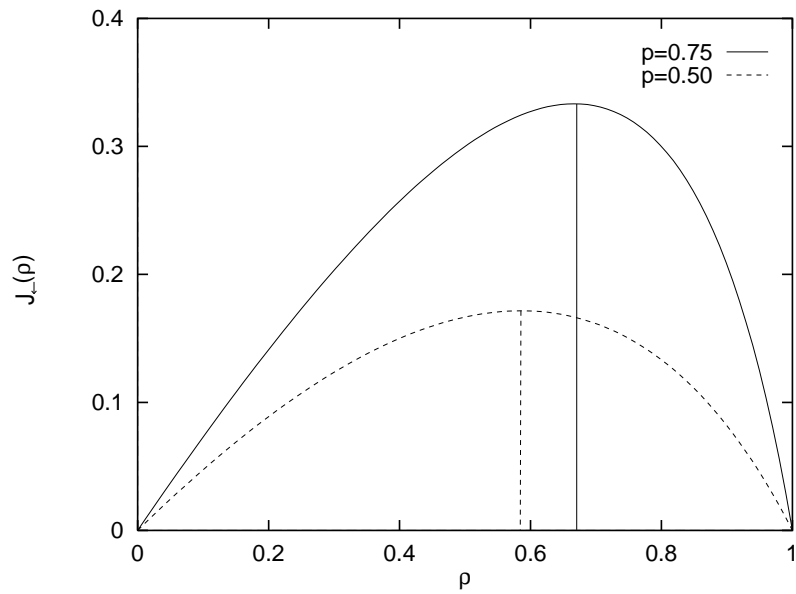,bbllx=10pt,bblly=10pt,bburx=550pt,bbury=440pt,
height=10cm}}
\caption{Fundamental diagram (flow vs.\ density) for the ASEP with 
ordered-sequential update and periodic boundary conditions. The vertical
lines indicate the location of $\rho^{\rm  max}_\leftarrow(p)$.}
\label{X0}
\end{figure}
\begin{figure}[t]
\centerline{\psfig{figure=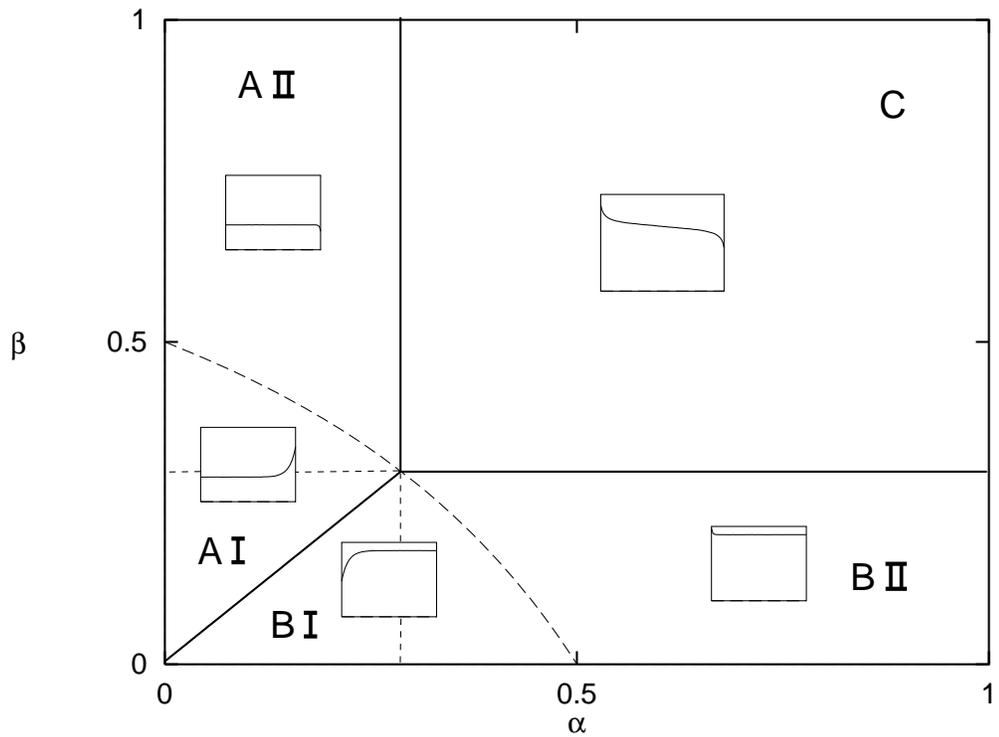,bbllx=10pt,bblly=10pt,bburx=570pt,bbury=420
pt,height=12cm}}
\caption{Phase diagram for the ASEP with ordered-sequential update 
$T_{\leftarrow}$ for $p=0.5$. The mean-field line (\ref{mean}) is curved 
dashed line. The straight dashed lines are the boundaries between the
phase AI and AII (BI and BII). The inserts show density profiles obtained 
from Monte Carlo simulations.}
\label{X1}
\end{figure}
\begin{figure}[p]
\newpage
\centerline{\psfig{figure=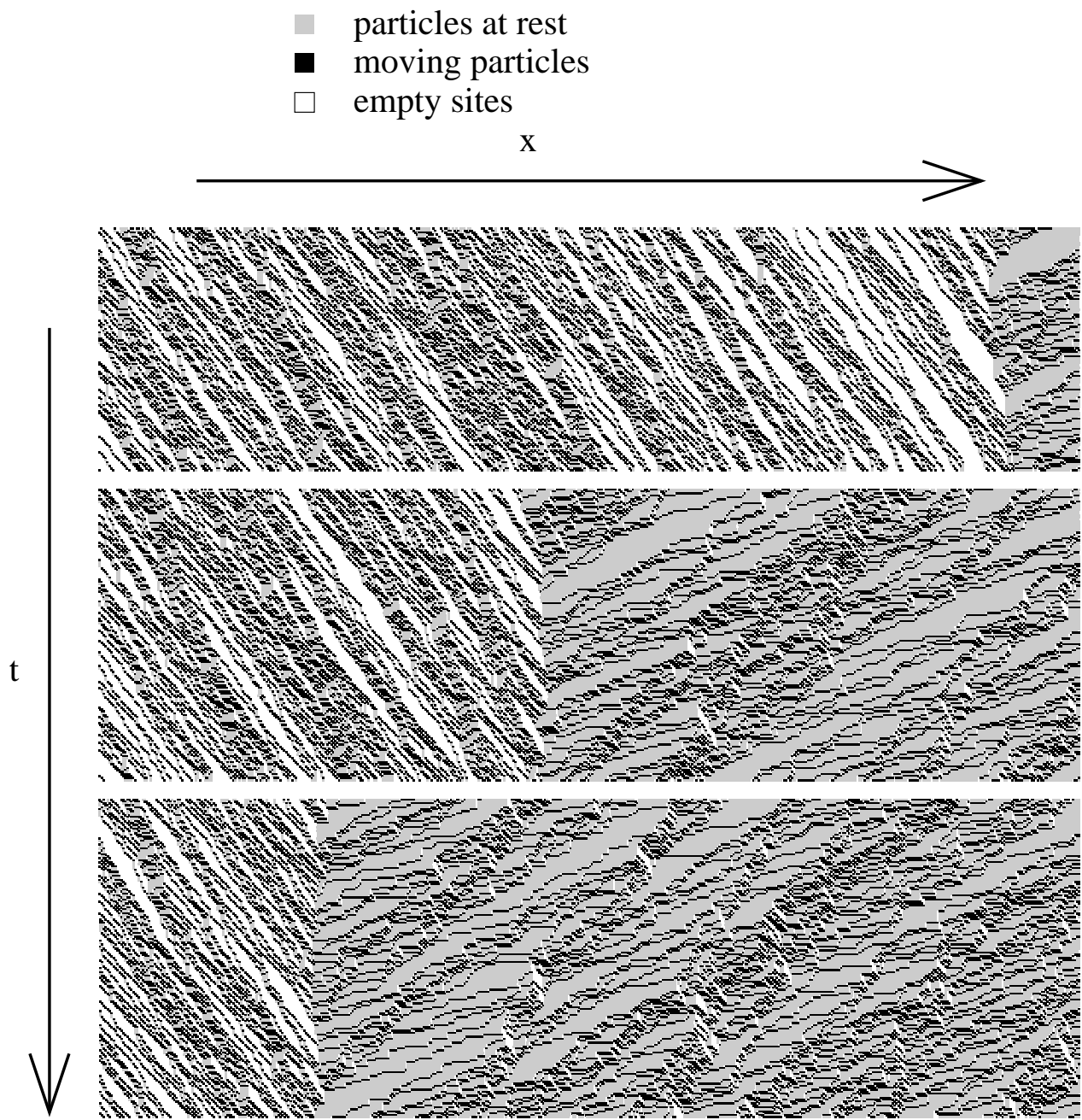,bbllx=10pt,bblly=40pt,bburx=580pt,bbury=730pt,height=21cm}}
\caption{The space-time diagram for
  $p=0.75,\alpha=\beta=0.3$. The microscopic shock moves freely on the
  chain.}
\label{X3}
\end{figure}
\begin{figure}[p]
\newpage
\centerline{\psfig{figure=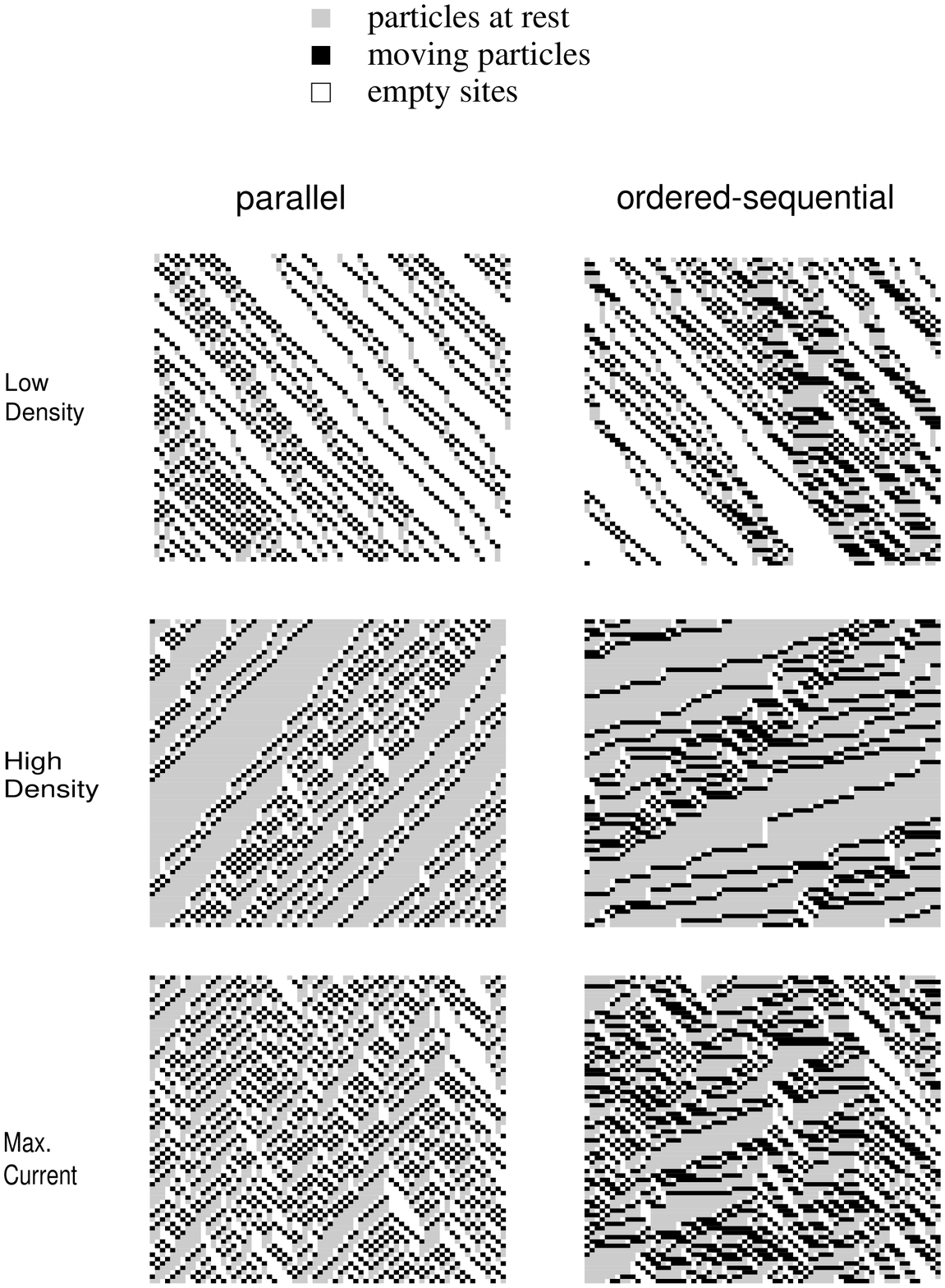,bbllx=10pt,bblly=40pt,bburx=580pt,
bbury=780pt,height=23cm}}
\caption{Typical space-time diagrams for the various phases.}
\label{X4}
\end{figure}
\begin{figure}[ht]
  \centerline{\psfig{figure=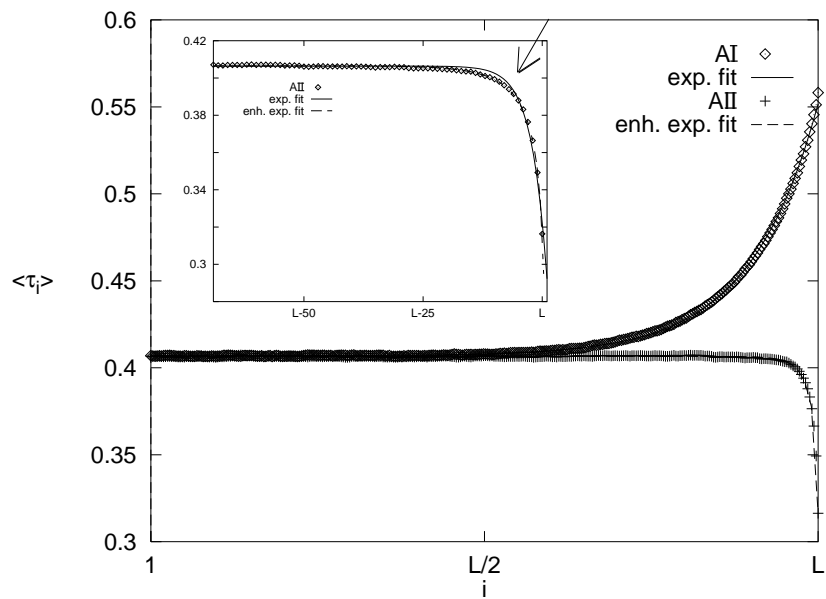,bbllx=50pt,bblly=20pt,bburx=570pt,bbury=415pt,height=10cm}}
  \caption{\protect{Density profiles for parallel update (p=0.75) in the 
      low-density phases AI and AII. The insert compares the  exact
      asymptotic form with a pure exponential decay. }\hfill}
\label{a12prof}
\end{figure}
\begin{figure}[ht]
  \centerline{\psfig{figure=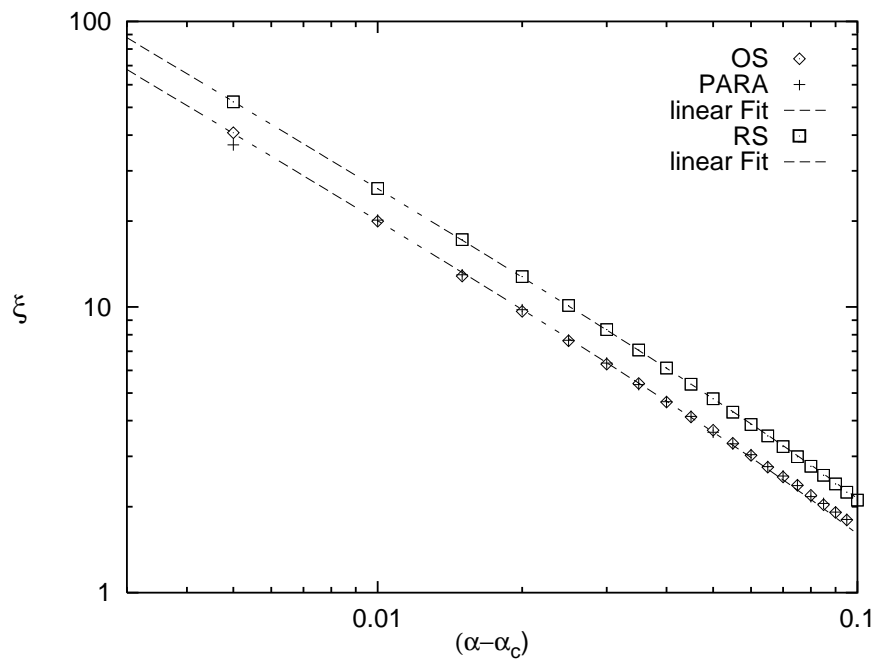,bbllx=60pt,bblly=10pt,bburx=570pt,bbury=420pt,height=10cm}}
  \caption{\protect{Log-Log plot of the correlation length for
      different types of updates (OS=ordered-sequential, PARA=parallel,
      RS=random-sequential). The length scales are obtained from
      an exponential fit of the density profile, $\alpha$  and $\beta$
      are chosen such that $\alpha + \beta =0.6$ holds
      $(\alpha_c=0.3)$. }\hfill}
\label{a1length}
\end{figure}
\begin{figure}[ht]
  \centerline{\psfig{figure=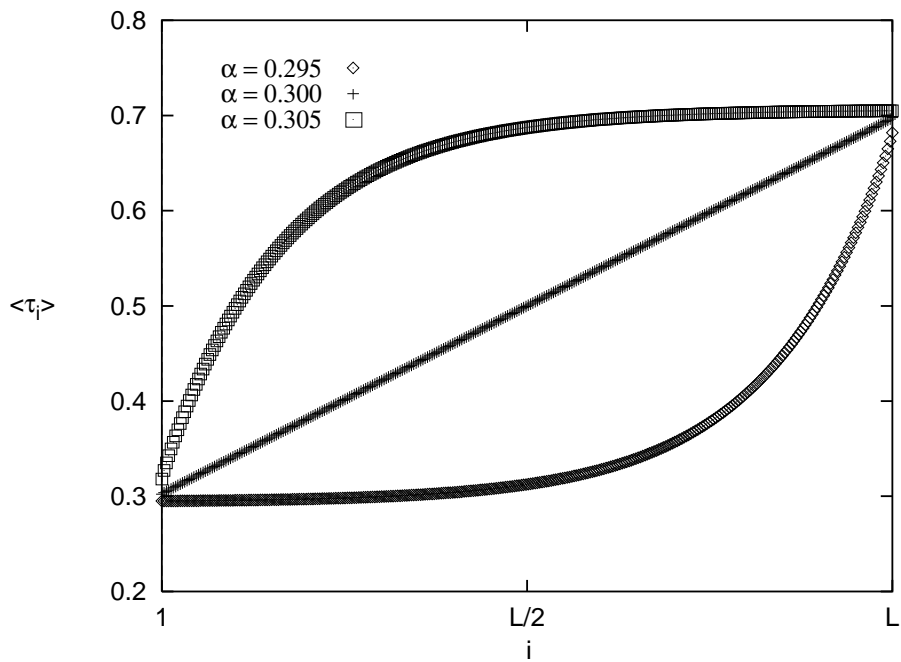,bbllx=35pt,bblly=20pt,bburx=570pt,bbury=430pt,height=10cm}}
  \caption{\protect{Density profiles near the first order transition
      at $\alpha  = \beta = 0.3$ for the random sequential
      update using the exact results of \cite{derrida93}. Again
      $\alpha + \beta =0.6$ holds. } \hfill}
\label{a1b1trans}
\end{figure}
\begin{figure}[ht]
  \centerline{\psfig{figure=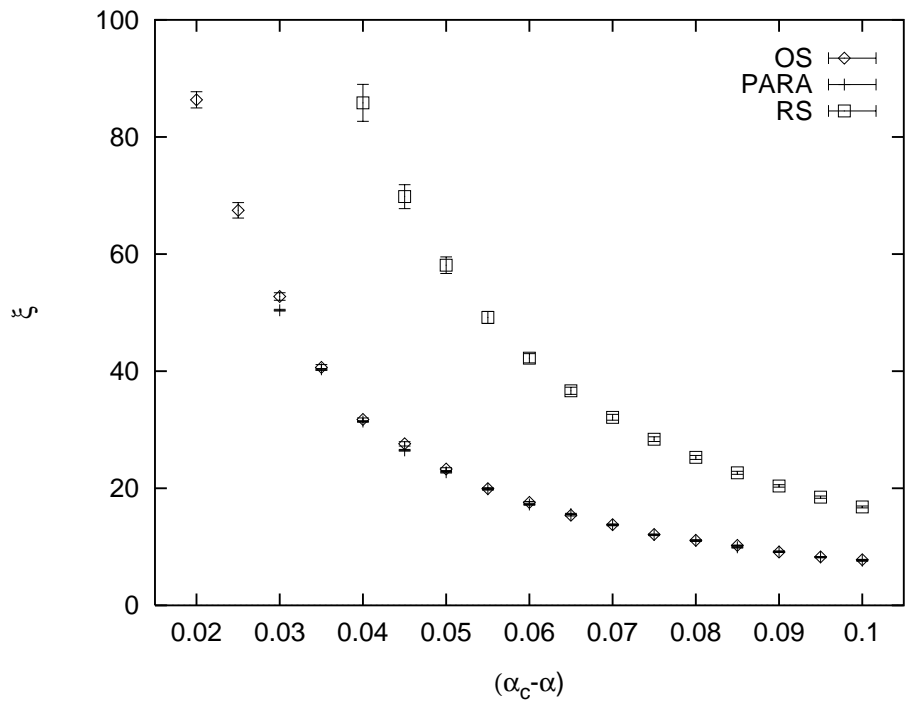,bbllx=50pt,bblly=20pt,bburx=570pt,bbury=420pt,height=10cm}}
  \caption{\protect{Divergence of the correlation length $\xi_\alpha$
      near the transition to the maximum current phase ($\beta =
 0.75$). }\hfill}
\label{a2length}
\end{figure}
\begin{figure}[ht]
  \centerline{\psfig{figure=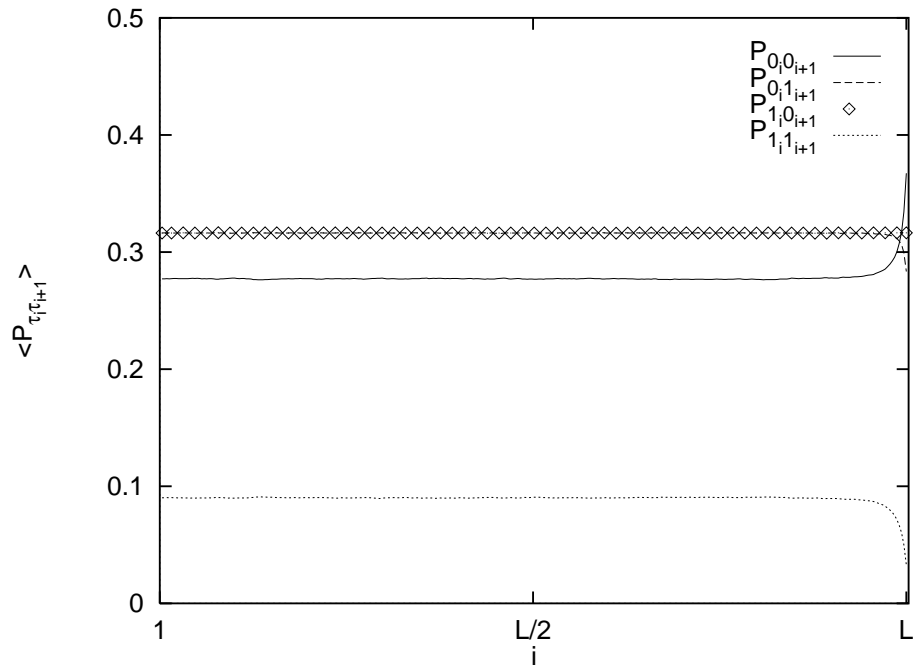,bbllx=30pt,bblly=20pt,bburx=570pt,bbury=420pt,height=10cm}}
  \caption{\protect{Density profile of the pair probabilities
      $P_{\tau_i\tau_{i+1}}$ for the parallel update in the low
      density phase AII ($\alpha = 0.40, \beta = 0.75, p = 0.75$).  }\hfill}
\label{pairs}
\end{figure}

\end{document}